\documentclass[aps, twocolumn, superscriptaddress]{revtex4} 

\usepackage[switch]{lineno}
\usepackage{amsmath}
\usepackage{amssymb}
\usepackage{empheq}
\usepackage[parfill]{parskip}
\usepackage[active]{srcltx}
\usepackage{color}
\usepackage{array}
\usepackage{booktabs}
\usepackage{amsfonts}
\usepackage{dsfont}
\usepackage{graphicx}

\begin{document}

\setlength{\parindent}{0.5cm}

\title{Collective behaviour of swarmalators on a 1D ring}







\author{Kevin O'Keeffe}
\affiliation{Senseable City Lab, Massachusetts Institute of Technology, Cambridge, MA 02139} 

\author{Steven Ceron}
\affiliation{Sibley School of Mechanical and Aerospace Engineering, Cornell University, Ithaca, NY 14853, USA}

\author{Kirstin Petersen}
\affiliation{Department of Electrical and Computer Engineering, Cornell University, Ithaca, NY 14853, USA}

\begin{abstract}
We study the collective behavior of swarmalators, generalizations of phase oscillators that both sync and swarm, confined to move on a 1D ring. This simple model captures some of the essence of movement in 2D or 3D but has the benefit of being solvable: most of the collective states and their bifurcations can be specified exactly. The model also captures the behavior of real-world swarmalators which swarm in quasi-1D rings such as bordertaxic sperm and vinegar eels. 
\end{abstract}

\maketitle

\section{Introduction}
Synchronization and swarming are universal phenomena \cite{winfree2001geometry,kuramoto2003chemical,pikovsky2003synchronization,toner1995long,vicsek1995novel} that are in a sense spatiotemporal opposites. Synchronizing units self-organize in time, but not in space; laser arrays fire simultaneously \cite{jiang1993numerical,kozyreff2000global}, heart cells trigger all at once \cite{peskin75}, but neither system has a notion of spontaneous group movement. Swarming units flip the picture: they self-organize in space, not time. Birds fly in flocks \cite{bialek2012statistical}, fish swim in schools \cite{katz2011inferring}, but neither coordinates the timing of an internal state or rhythm. 

The units of some systems appear to self-organize in both space and time. In Biology, sperm \cite{yang2008cooperation,riedel2005self}, vinegar eels \cite{quillen2021metachronal,quillen2021synchronized}, and other microswimmers \cite{taylor1951analysis,tamm1975role,huang2021circular} synchronize the wriggling of their tails which is speculated to hydrodynamically couple to their motion. In Chemistry, magnetic Janus particles \cite{yan2012linking,yan2015rotating,hwang2020cooperative}, dieletric Quinke rollers \cite{zhang2020reconfigurable,bricard2015emergent,zhang2021persistence}, and other active entities \cite{manna2021chemical} lock their rotations enabling a kind of sync-dependent self-assembly. In Engineering, land-based robots and aerial drones can be programmed to swarm based on their synchronizable internal clocks \cite{barcis2019robots,barcis2020sandsbots,monaco2020cognitive}. Sync and swarming are also suspected to interact in spatial cognition \cite{monaco2019cognitive}, embryology \cite{uriu2017determining,uriu2014collective,tsiairis2016self} and the physics of magnetic domain walls \cite{hrabec2018velocity}.

The theoretical study of systems which both sync and swarm is just beginning. Takana pioneered the initiative by formulating a model of chemotactic oscillators \cite{tanaka2007general} and found diverse phenomena \cite{iwasa2017mechanism,iwasa2010dimensionality}. Leibchen and Levis generalized the Vicsek model and found new types of long range synchrony \cite{levis2019activity}. O'Keeffe et al introduced a model of `swarmalators' \footnote{Short for `swarming oscillators'.} whose collective states have been realized in Nature and technology \cite{barcis2019robots,barcis2020sandsbots,zhang2020reconfigurable} and is being actively extended. The inclusion of noise \cite{hong2018active}, local coupling \cite{lee2021collective,jimenez2020oscillatory}, periodic forcing \cite{lizarraga2020synchronization}, mixed sign interactions \cite{mclennan2020emergent}, and finite $N$ effects \cite{o2018ring} have been studied. The potential of swarmalators in bio-inspired computing has been explored \cite{o2019review}, as has the well-posedness of $N \rightarrow \infty$ solutions of the swarmalator model \cite{ha2021mean,ha2019emergent}.

Here we study swarmalators confined to move on a 1D ring. Our aim is two-fold. First, to model swarmalators which swarm purely in 1D. Frogs, nematodes, and other organisms are often bordertaxic, seeking out the ring-like edges of their confining geometry \cite{bau2015worms,yuan2015hydrodynamic,nosrati2016predominance,ketzetzi2021activity,creppy2016symmetry,aihara2009modeling,aihara2014spatio}. Janus particles, when acting as microrobots for precision medicine \cite{le2019janus,agrawal2019janus,yi2016janus}, will need to navigate pseudo 1D grooves and channels \cite{xiao2018review}. Toy models for these systems will be useful for applied research.


\begin{figure}
    \centering
    \includegraphics[width= \columnwidth]{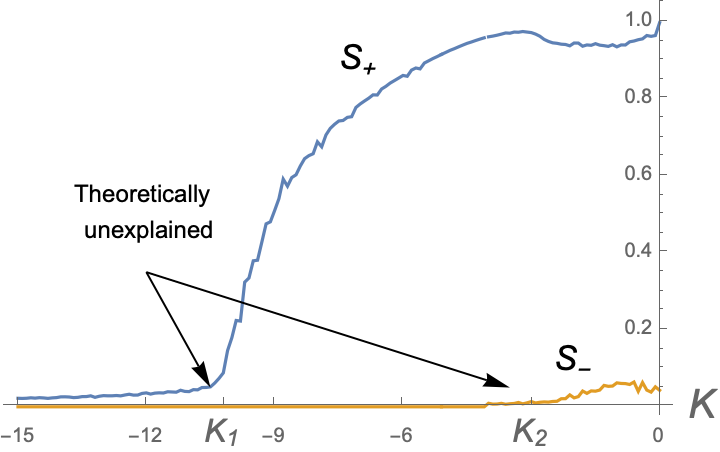}  
    \caption{Order parameters of the 2D swaramalator model $S_{\pm} e^{i \phi_{\pm}} := (N)^{-1} \sum_j e^{i(\phi_j \pm \theta_j)}$, where $\phi, \theta$ are the spatial angle and phase of swarmalators. $S_+$ bifurcates from 0 at $K_1$ as the static async state (Fig~\ref{states-2D}(b)) destabilizes, $S_-$ from $0$ at $K_2$ as the active phase wave (Fig~\ref{states-2D}(f)) destabilizes. The nature of these bifurcations, as well as analytic expressions for both $K_1$ and $K_2$, are unknown.} 
    \label{order-parameters-2D}
\end{figure}

\begin{figure*}
    \centering
    \includegraphics[width= 2.0
    \columnwidth]{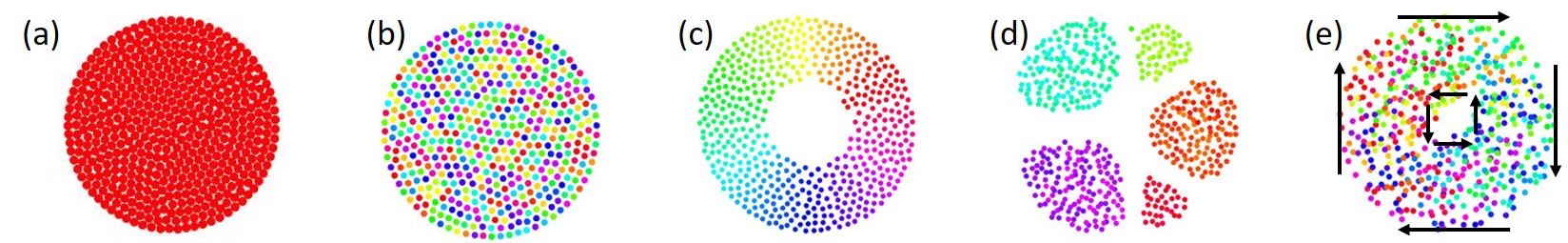}
    \caption{Collective states of the 2D swarmalator model introduced in \cite{o2017oscillators} where swarmalators are represented as colored dots where the color refers to the swarmalators phase. As described in Appendix B, the states displayed are from a slightly different intstance of the model to that presented in \cite{o2017oscillators} which produces the same qualitative behavior, but better emphasises the non-monotonic behavior of the order parameters $S{\pm}$ shown in Figure~\ref{order-parameters-2D}. In all panels a Euler method was used with timestep $dt = 0.1 $ for $T = 1000$ units for $N = 1000$ swarmalators. (a) Static sync: $(J,K,\sigma) = (1,1,10)$ (b) Static async $(J,K,\sigma) = (1,1,10)$ (c) Static phase wave $(J,K,\sigma) = (1,1,10)$ (d) Splintered phase wave $(J,K,\sigma) = (1,1,10)$ (e) Active phase wave. In the three static states (a)-(c) swarmalators do not move in space or phase. In the splintered phase wave, each colored chunk is a vortex: the swarmalators librate in both space and phase. In the active phase wave, the librations are excited into rotations; the swarmalators split into counter-rotating groups as indicated by the black arrows. } 
    \label{states-2D}
\end{figure*}

Our second aim is theoretical, namely to investigate the original 2D swarmalator model whose physics is not understood \footnote{Note the Figure displays the order parameters of the instance of the model; see Appendix B}. Figure~\ref{order-parameters-2D} shows its order parameters $S_{\pm} := (N)^{-1} \sum_j e^{i(\phi_j \pm \theta_j)}$ -- where $\phi, \theta$ denote the spatial angle and phase -- dependence on the coupling strength $K$. At a critical $K_1$, $S_+$ jumps from zero as the system transitions from a static async state (Fig~\ref{states-2D}(b)) to an active phase wave state (Fig~\ref{states-2D}(e)) in which swarmalators run in a space-phase vortex. As expected of order parameters, $S_+$ grows as $K$ is increased. But then it begins to decline at a second value $K_2$ as the swarmalator vortex bifurcates into a broken band of mini-vortices (Fig~\ref{states-2D}(e)). The cause of this non-monotinicity, as well as analytic values for $K_1, K_2$, are unknown. Like the old puzzles to understand the transitions of the Kuramoto model \cite{kuramoto2003chemical,strogatz2000kuramoto,strogatz1991stability,mirollo2007spectrum,crawford1994amplitude}, the bifurcations of the swarmalator model `cry out for a theoretical explanation' \cite{strogatz2000kuramoto}.


Our hope is that a retreat to a 1D ring will give some first clues on how to provide such an explanation.


\section{Model}
We study a pair of modified Kuramoto models,
\begin{align}
    \dot{x_i} &= \nu_i + \frac{J}{N} \sum_j^N \sin(x_j - x_i) \cos(\theta_j - \theta_i) \label{eom-x} \\
    \dot{\theta_i} &=  \omega_i + \frac{K}{N}  \sum_j^N \sin(\theta_j - \theta_i ) \cos(x_j - x_i ) \label{eom-theta}
\end{align}
\noindent
where $(x_i, \theta_i) \in (S^1, S^1)$ are the position and phase of the $i$-th swarmalator for $i = 1, \dots, N$ and ($\nu_i, \omega_i$), $(J,K)$ are the associated natural frequencies and couplings. We consider identical swarmalators $(\omega_i, \nu_i) = (\omega, \nu)$ and by a change of frame set $\omega = \nu = 0$ WLOG.

Eq.~\eqref{eom-theta} models position-dependent synchronization. The familiar Kuramoto sine term minimizes swarmalators' pairwise phase difference (so they sync) while the new cosine term strengthens the coupling between nearby swarmalators $K_{ij} = K \cos(x_j - x_i)$ (so the sync is position dependent). Eq.~\eqref{eom-x} is Eq.~\eqref{eom-theta}'s mirror-image: it models phase-dependent swarming. Now the sine term minimizes swarmalators' pairwise distances (so they swarm / aggregate) and the cosine term strengthens the coupling between similarly phased swarmalators $J_{ij} = J \cos(\theta_j - \theta_i)$. You can also think of Eqs.\eqref{eom-x} and \eqref{eom-theta} as modelling synchronization on the unit torus.

Converting the trig functions to complex exponentials and rearranging makes the model even simpler
\begin{align}
    \dot{x_i} &= \frac{J}{2} \Big(  S_+ \sin( \Phi_+ - (x_i + \theta_i) ) + S_- \sin( \Phi_- - (x_i - \theta_i) )  \Big) \\
    \dot{\theta_i} &= \frac{K}{2} \Big(  S_+ \sin( \Phi_+ - (x_i + \theta_i) ) - S_- \sin( \Phi_- - (x_i - \theta_i) )  \Big)
\end{align}
\noindent
where 
\begin{equation}
    W_{\pm} = S_{\pm} e^{i \Phi_{\pm}} =  \frac{1}{N} \sum_j e^{i(x_j \pm \theta_i)}. \label{order-par}
\end{equation}
The terms $x_i \pm \theta_i$ occur naturally so we define 
\begin{align}
    \xi_i = x_i + \theta_i \label{Wp} \\
    \eta_i = x_i - \theta_i \label{Wm}
\end{align}
And find
\begin{align}
    \dot{\xi_i} &=  J_+ S_+ \sin( \Phi_+ - \xi ) + J_{-} S_{-} \sin( \Phi_- - \eta ) \label{eom_xi} \\
    \dot{\eta_i} &=  J_+ S_+ \sin( \Phi_+ - \xi ) + J_+ S_- \sin( \Phi_- - \eta ) \label{eom_eta} 
\end{align}
\noindent
where $J_{\pm} = (J \pm K) / 2$ \footnote{Note we could set $J = 1$ WLOG by rescaling time but we decided against this so as to make the facilate a clean comparison to the 2D model for which both $(J,K)$ appear}.

We see ring swarmalators obey a sum of two Kuramoto models where the traditional order parameter $R e^{i \Phi} := (N)^{-1} \sum_j e^{i \theta_j}$ has been replaced by a pair of new order parameters $W_{\pm} = S_{\pm} e^{i \Phi_{\pm}} = (N)^{-1} \sum_j e^{i (x_j \pm \theta_j)}$. The new $W_{\pm} $ measure the system's total amount of `space-phase order`. 

\begin{figure*}
    \centering
    \includegraphics[width= 2.0 \columnwidth]{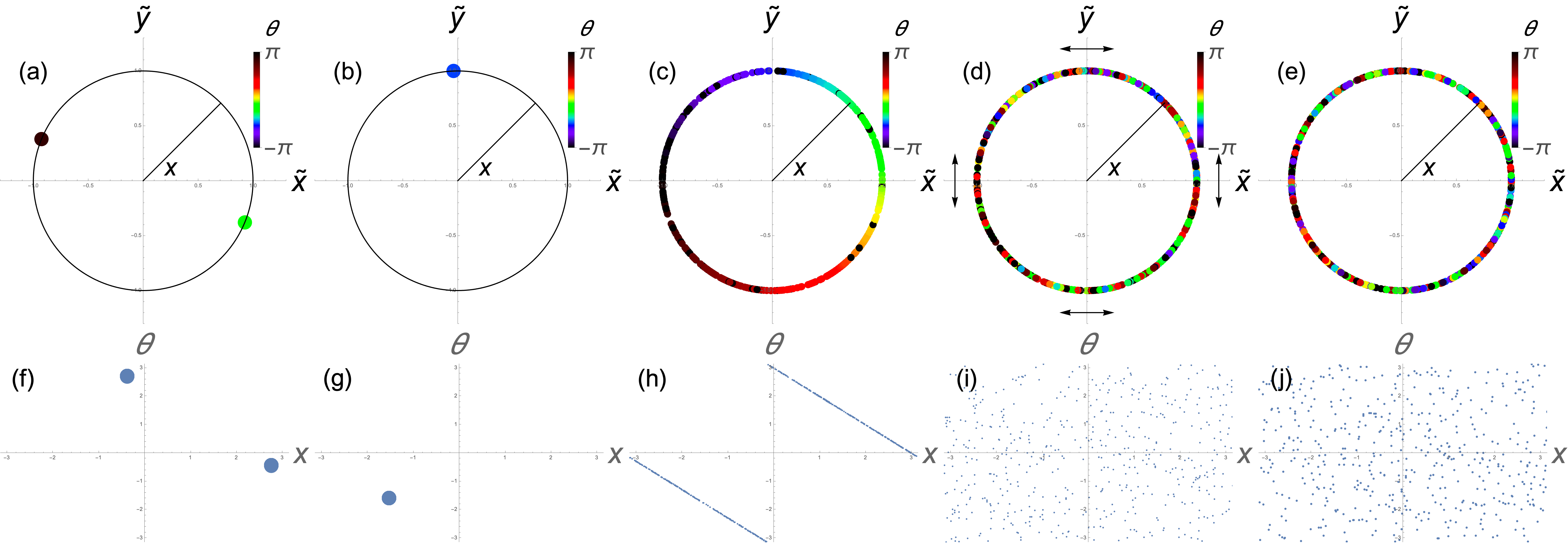}
    \caption{Collective states of ring model. Top row, swarmalators are drawn in as colored points on a unit circle in a dummy $(\tilde{x}, \tilde{y})$ plane to allow a comparison with the 2D swarmalator model in Figure~\ref{states-2D}. $x$ corresponds to the angular position on this unit circle (drawn black in (a) and (b) for greater clarity. The point sizes in these panels, and the corresponding panels (f),(g) are also larger, again to make things clearer) while the color corresponds to the phase. Bottom row, swarmalators are drawn as points in the $(x, \theta)$ plane. All results were found by integrating Eqs.~\eqref{eom-x}, \eqref{eom-theta} using an RK4 solver with timestep $dt = 0.1$ for $T = 500$ time units for $N = 500$ swarmalators. Initial positions and phases were drawn from $[-\pi, \pi]$ in all panels except panel (a), which were drawn from $[0,\pi]$ (since this choice of initial conditions realized the static sync state). (a),(f) Static sync $(J,K)=(1,1)$ (b),(g) Static $\pi$-state $(J,K)=(1,1)$ (c),(h) Static phase wave $(J,K)=(1,-0.5)$ (d),(i) Active async $(J,K) = (1,-1.05)$. Here the swarmalators jiggle about in $(x,\theta)$ as indicated by the double ended arrows with no global space-phase order as indicated by the scatter plot in (i). The amount of motion / jiggling depends on the population size $N$ as discussed in the main text and illustrated in Figure~\ref{v-N-K}. (e), (j) Static async $(J,K) = (1,-2)$.}
    \label{states-1D}
\end{figure*}

What kind of order is this? The limiting cases are trivial: static sync, $(x_i, \theta_i) = (x^*, \theta^*)$, which produces maximal order $S_{\pm} = 1$ (which follows from substitution into Eq.\eqref{order-par}), and static async, in which positions $x_i$ are fully uncorrelated with phases $\theta_i$, which produces minimal order $S_{\pm} = 0$. Between these two extremes, however, lies something more interesting: Perfect correlation between position and phase $x_i = \pm \theta_i + c$ for constant $c$ which yields either $(S_+, S_-) = (0,1)$ or $(S_+, S_-) = (0,1)$ \footnote{If the correlation is $x_i + \theta_i = c $, then $S_+ = 1$. If $x_i - = \theta_i + c$, then $S_- = 1$.}. What does $x_i \pm \theta_i$ mean physically? Picture swarmalators on a ring as a group of fireflies flying around a circular track, flashing periodically. Let $\theta_i = 0$ be the beginning of their phase cycle (when they flash). Then $x_i \pm \theta_i +c $ means each firefly flashes at the same point on its circular lap $x_i = c$. Plotting swarmalators as colored dots in space (where color represents phase) illustrates this behavior most evocatively. Then $x_i = \pm \theta_i + c$ corresponds to a colored splay state (Fig~\ref{states-2D}(c), Fig~\ref{states-1D}(c)). Since one of $S_{\pm}$ are maximal in this rainbow-like state, we call $S_{\pm}$ the \textit{rainbow order parameters}. 

As we will show, $S_{\pm}$ are natural order parameters for the ring model (insofar as they can distinguish between each of its collective states).

\newpage
\textbf{Motivation for model}. Before showing our analysis, we quickly show how the ring model connects to the original 2D model \cite{o2017oscillators}. The 2D model is
\begin{align}
&\dot{\mathbf{x}}_i = \mathbf{v}_i + \frac{1}{N} \sum_{j=1}^N \Big[ \mathbf{I}_{\mathrm{att}}(\mathbf{x}_j - \mathbf{x}_i)F(\theta_j - \theta_i)  - \mathbf{I}_{\mathrm{rep}} (\mathbf{x}_j - \mathbf{x}_i) \Big],  \label{x_eom_model} \\ 
& \dot{\theta_i} = \omega_i +  \frac{K}{N} \sum_{j=1}^N H_{\mathrm{\mathrm{att}}}(\theta_j - \theta_i) G(\mathbf{x}_j - \mathbf{x}_i) \label{theta_eom_model}
\end{align}
We pick out the rotational component of the swarming motion -- since that part is analogous to movement on a ring -- by converting Eqs.\eqref{x_eom_model}, \eqref{theta_eom_model} to polar coordinates. For certain choices of $\mathbf{I}_{\mathrm{att}}(\mathbf{x}), F(\theta)$ etc this yields
\begin{align}
\dot{r_i}&= \tilde{\nu}(r_i) +   \frac{J}{2} \Bigg[ \tilde{S}_+ \cos \Big( \Phi_+- \xi_i \Big) + \tilde{S}_{-}\cos \Big( \Phi_{-} - \eta_i \Big) \Bigg] \label{2D_radial} \\
\dot{\xi_i}&= \tilde{\omega}(r_i) + \tilde{J}_+(r_i) \tilde{S}_+ \sin \Big( \Psi_+ - \xi_i \Big) \nonumber \\ & \qquad \qquad  + \tilde{J}_- (r_i) \tilde{S}_- \sin \Big( \Psi_{-} -  \eta_i \Big) \label{t1} \\
\dot{\eta_i} &= \tilde{\omega}(r_i) +  \tilde{J}_-(r_i) \tilde{S}_+ \sin \Big( \Psi_+ - \xi_i \Big) \nonumber \\ & \qquad \qquad + \tilde{J}_+(r_i) \tilde{S}_- \sin \Big( \Psi_- - \eta_i \Big) \label{t2}
\end{align}
where we have switched to $(\xi_i, \eta_i) = (\phi_i + \theta_i, \phi_i - \theta_i)$ coordinates and $\phi_i$ is the spatial angle of the $i$-th swarmalator (which is analogous to $x_i$ in the ring model). We put the derivation and definitions of the various new quantities in the Appendix because they are cumbersome and uninformative to display here.

Eqs.~\eqref{t1}, \eqref{t2} reveal the ring model hidden in the 2D model's core. The $(\dot{\xi}_i, \dot{\eta}_i)$ equations have the same form as the $(\dot{\xi_i}, \dot{\eta_i})$ equations in the 1D model (Eqs.~\eqref{eom_xi}, \eqref{eom_eta}); both are a summed pair of Kuramoto models. The only difference is that in the 2D model, $\tilde{\omega}_i(r_i)$ and $\tilde{J}_{\pm}(r)$ depend on $r_i$. So the ring model is like the 2D model with the radial dynamics turned off. This is why we think studying the ring model will yield hints on how to study the 2D model (one of the paper's aims).



\section{Results}

\begin{figure}
    \centering
    \includegraphics[width= \columnwidth]{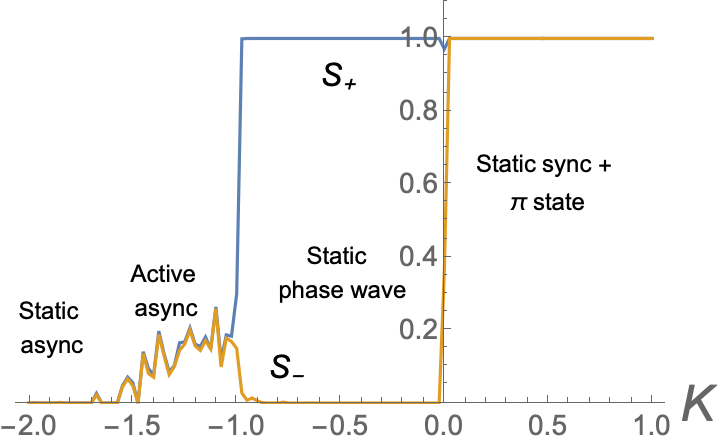}
    \caption{Rainbow order parameters $S_{\pm}$ for the ring swarmalator model with $\nu_i = \omega_i = 0$ and $J = 1$. We have assumed WLOG that $S_+ > S_i$ (i.e. if $S_+ < S_-$ in simulations we swapped $(S_+, S_-) \rightarrow (S_-, S_+)$). Data were collected by integrating the model \eqref{eom-x}, \eqref{eom-theta} using an RK4 method with $(dt, T) = (0.1, 500)$. The first $90 \% $ of data were discarded as transients and the mean of the remaining $10\%$ were taken. We chose just $N = 10$ swarmalators to illustrate the active async state as clearly as possible; the fluctuations in $S_{\pm}$ that characterize the state decay to $0$ for larger $N$ (see main text).}
    \label{order-parameters-1D-identical}
\end{figure}

\subsection{Numerics}

Simulations show the system settles into five collective states depicted in Figure~\ref{states-1D}. We provide the Mathematica code used for the simulations at \footnote{https://github.com/Khev/swarmalators/blob/master/1D/on-ring/regular/sandbox.nb}.

\begin{enumerate}
    \item \textit{Static sync} (Fig~\ref{states-1D}(a),(f)). Swarmalators fully synchronize their positions $x_i = x^*$ and phases $\theta_i = \theta^*$ resulting in maximal space-phase order $S_{\pm} = 1$.
    \item \textit{Static $\pi$ state} (Fig~\ref{states-1D}(b),(g)). One group of swarmalators synchronize at $(x^*, \theta^*)$ and the remaining fraction synchronize $\pi$ units away $(x^* + \pi, \theta^* + \pi)$. $S_{\pm} = 1$ here also. 
    \item \textit{Static phase wave} (Fig~\ref{states-1D}(c),(h)). Swarmalators form a static splay state with $x_i = 2 \pi i / N$ and $\theta_i = \pm x_i + c$ where the offset $c$ is arbitrary and stems from the rotational symmetry in the model \footnote{Linear transformations $x_i \rightarrow x_i + C$ and $\theta_i \rightarrow C$ do not change the dynamics because only differences $x_j - x_i$ and $\theta_j - \theta_i$ appear in the model.}. In $(\xi_i, \eta_i)$ coordinates, either $\xi_i$ is splayed $\xi_i = 2\pi i /N$ and $\eta_i$ is locked $\eta_i = c$ or vice versa. The order parameters are either $(S_+, S_-) = (1,0)$ (where phase gradient of the rainbow is clockwise) or $(S_+, S_-) = (1,0)$ (where the phase gradient of the rainbow is counter-clockwise).
    \item \textit{Active async} (Fig~\ref{states-1D}(d),(i)). Swarmalators form a dynamic steady state, moving in clean limit cycles for small $N$, but in erratic, jiggling patterns for large $N$. The motion cools and ultimately freezes as $N \rightarrow \infty$. There is little space-phase order as indicated by the low, time averaged values of $S_{\pm}$ (Fig~\ref{order-parameters-2D}) so we call this state 'active async'. 
    \item \textit{Static async} ((Fig~\ref{states-1D}(e),(j))). Swarmalators form a static, asynchronous crystal with $S_{\pm} = 0$.
\end{enumerate}

Figure~\ref{order-parameters-1D-identical} shows the curve $S_\pm(K)$ can distinguish between all but the static sync and static $\pi$ states.

Now we analyze the stability of the states. The static sync, $\pi$, and static async states are analyzed using standard techniques, but are useful as warm ups to the analysis of the much harder static phase wave state (which is the main analytic contribution of the paper). The active async state, being non-stationary, is analyzed mostly numerically.

\subsection{Analysis}
\textbf{Static sync}. We calculate the stability of the state by linearizing around the fixed point in $(\xi, \eta)$ space. We seek the eigenvalues $\lambda$ of the Jacobian $M$
\begin{equation}
    M = \left[ 
\begin{array}{cc} 
  Z_{\xi} & Z_{\eta} \\ 
  N_{\xi} & N_{\eta} \\
\end{array} 
\right] 
\end{equation}
where
\begin{align}
(Z_{\xi})_{ij} &= \frac{\partial \dot{\xi}_i}{\partial \xi_j} \\
(Z_{\eta})_{ij} &= \frac{\partial \dot{\xi}_i}{\partial \eta_j} \\
(N_{\xi})_{ij} &= \frac{\partial \dot{\eta}_i}{\partial \xi_j} \\
(N_{\eta})_{ij} &= \frac{\partial \dot{\eta}_i}{\partial \eta_j} 
\end{align}
Evaluating the derivatives in the above using Eqs.~\eqref{eom_xi}, \eqref{eom_eta} results in a clean block structure:
\begin{equation}
    M = \left[ 
\begin{array}{cc} 
  J_+ A(\xi) & J_- A(\eta) \\ 
  J_- A(\xi) & J_+ A(\eta) \\
\end{array} 
\right] 
\end{equation}
where $A(y)_{i,i} = - \frac{1}{n} \sum_{j=1}^n \cos(y_j - y_i)$ and $A(y)_{i,j} = \frac{1}{n} \cos(y_j - y_i)$ for a dummy variable $y$. Evaluated at the fixed points of the static sync state, $\xi_i = c_1$ and $\eta_i = c_2$ for constants $c_1, c_2$, this becomes even simpler,
\begin{equation}
    M_{SS} = \left[ 
\begin{array}{cc} 
  J_+ A_0 & J_- A_0 \\ 
  J_- A_0 & J_+ A_0 \\
\end{array} 
\right] \label{Mss}
\end{equation}
where
\begin{equation}
A_0 := \begin{bmatrix}
    - \frac{N-1}{N}   & \frac{1}{N}  & \dots & \frac{1}{N}\\
    \frac{1}{N}  & - \frac{N-1}{N} &  \dots & \frac{1}{N} \\
    \hdotsfor{4} \\
    \frac{1}{N}  & \frac{1}{N}  & \dots & - \frac{N-1}{N} 
\end{bmatrix} \label{A0}
\end{equation}
Notice that Jacobian of the entire system $M$ has $\dim(N) = 2N$ since there are two state variables $(x, \theta)$ for each of the $N$ swarmalators, but that $\dim(A_0) = N$ since it is a subblock of $M$. Now, the eigenvalues $\hat{\lambda}$ of $A_0$ are well known: there is $1$ with value $\hat{\lambda} = 0$ stemming from the rotational symmetry of the model and $N-1$ $\hat{\lambda} = -1$. We use these to find the desired eigenvalues $\lambda$ of $M_{SS}$ using the following identity for symmetric block matrices 
\begin{equation}
    \det E := \det \left[ 
\begin{array}{cc} 
  C & D \\ 
  D & C \\
\end{array} 
\right] = \det(C+D) \det(C-D)
\end{equation}
This implies the eigenvalues of $E$ are the union of the eigenvalues of $C + D$ and $C-D$. Applying this identity to $M_{SS}$ (which has the required symmetric structure) yields
\begin{align}
    \lambda_0 &= 0 \\
    \lambda_1 &= - J \\
    \lambda_2 &= -K
\end{align}
with multiplicities $2, -1 + N, -1 + N$ (which sum to the required $2N$). This tells us the static sync is stable for $J > 0$ and $K > 0$ consitent with simulations.

\textbf{Static $\pi$-state}. The fixed points here are $(x_i, \theta_i) = (c_1, c_2)$ for $i = 1,2, \dots, N/2$ and $(x_i, \theta_i) = (c_1 + \pi, c_2 + \pi)$ for $i = N/2, \dots, N$. Conveniently, the shift in $\pi$ for exactly half of swarmalators does not change the form of the Jacobian, $M_{\pi} = M_{SS}$, so the stability is the same as before. This means the static sync and $\pi$ states are bistable for all $J > 0, K > 0$. Appendix A discusses the basins of attraction for each state.

\textbf{Static phase wave}. We calculate the stability of the static phase wave using the same strategy as before: linearize around the fixed points and exploit the block structure of the Jacobian $M$. This time, however, the calculations are tougher.

The fixed points of the static phase wave take two forms: Either $\xi$ is splayed $\xi_i = 2 \pi (i-1) / N + c_1 $ and $\eta_i$ is synchronized $\eta_i = c_2$ (clockwise rainbow) or $\xi$ is synchronized $\xi_i = c_1$ and $\eta$ splayed $\eta_i = 2 \pi (i-1) / N +  c_2$ (counter clockwise rainbow). Here $i = 1, \dots, N$ and the constants $c_1, c_2$ are offsets. WLOG we analyze the fixed point with $\xi$ splayed and $\eta$ sync'd. The Jacobian is
\begin{equation}
    M_{SPW} = \left[ 
\begin{array}{cc} 
  J_+ A_1 & J_- A_0 \\ 
  J_- A_1 & J_+ A_0 \\
\end{array} 
\right]  \label{M_spw}
\end{equation}
where $A_0$ is as before Eq.\eqref{A0} but $A_1$ is new:
\begin{align}
    (A_1)_{ii} &= -\frac{1}{N} \sum_{j \neq i} \cos \frac{2\pi}{N} (j-i) := -\frac{1}{N} \sum_{j \neq i} c_{ij} \label{A1ii}\\
    (A_1)_{ij} &= \frac{1}{N}\cos \frac{2\pi}{N} (j-i) := \frac{1}{N} c_{ij} 
\end{align}
The following notation will be useful
\begin{align}
    c_{ij} & := \cos 2 \pi(i-j) / N \\
    s_{ij} & := \sin 2 \pi(i-j) / N \\
    \beta_{ij} & := c_{ij} + \mathcal{I} s_{ij} \\
    c_i & := \cos 2 \pi i / N \\
    s_i & := \sin 2 \pi i / N 
\end{align}
where $\mathcal{I} = \sqrt{-1}$ is the imaginary unit \footnote{We choose the non-standard notation $\mathcal{I}$ since we have already used $i$ and $j$ as indices.}, $\beta_{0,0} := \beta = e^{2\pi \mathcal{I} N}$ is the primitive root of unity, and $\beta_k = \beta^k$ is the $k$-th root of unity. The diagonal element $A_{ii}$ may be simplified. Recalling the sum of roots of unities are zero (you can think of $\beta_k$ as a vector pointing to the beginning of the $k$-th segment of size $1/N$ of the unit circle; then summing all the vectors around the unit circle results in zero)
\begin{equation}
    \sum_{k=0}^{N-1} \beta_k = \sum_{k=0}^{N-1} c_k + \mathcal{I} s_k = 0
\end{equation}
which implies
\begin{align}
    \sum_{k=0}^{N-1} c_k  = 0 \label{cos_zero} \\
    \sum_{k=0}^{N-1} s_k = 0 \label{sin_zero}
\end{align}
(In other words, the discrete sum of $\cos(2\pi k /N)$ and $\sin(2 \pi k N)$ around the unit circle is zero, which can be seen by symmetry). Applying these identities to Eq.~\eqref{A1ii} for $(A_1)_{ii}$,
\begin{align}
(A_1)_{ii} &= \frac{1}{N} \sum_{j \neq i} \cos \frac{2\pi}{N} (j-i) = -\frac{1}{N} \sum_{k=1}^{N-1} c_k \\
        &= -\frac{1}{N}( -c_0 + \frac{1}{N} \sum_{k=0}^{N-1} c_k) = -\frac{1}{N}( -c_0 + 0)  \\
        &=  \frac{c_0}{N}
\end{align}
Notice in the first sum over $i$ the $j-th$ term is excluded which means the second sum over $k$ begins at $1$. Since $c_0 = c_{0,0} = 1$, we get
\begin{align}
    (A_1)_{ij} = \frac{1}{N} c_{i,j} \label{A1}
\end{align}
With $A_0$ (Eq.~\eqref{A0}) and $A_1$ (Eq.~\eqref{A1}) in hand, we can begin finding the eigenvalues of $M_{SPW}$ Eq.~\eqref{M_spw} by using another identity for block matrices. If the consituent matrices $A,C$ of a general block matrix $P$,
\begin{equation}
    P = \left[ 
\begin{array}{cc} 
  A & B \\ 
  C & D \\
\end{array} 
\right]  
\end{equation}
commute then 
\begin{equation}
    \det(P) = \det(AD - BC) \label{commute}
\end{equation}
Luckily, the constituent matrices $A_0$ and $A_1$ of $M_{SPW}$ do commute. So we apply the above identity to the characteristic equation for $\lambda$
\begin{equation}
    \det(M_{SPW} - \lambda I) = \det \left[ 
\begin{array}{cc} 
  J_+ A_1 - \lambda I & J_- A_0 \\ 
  J_- A_1 & J_+ A_0 - \lambda I \\
\end{array} 
\right]  = 0
\end{equation}
And find
\begin{align}
\det(M_{SPW} - \lambda I) &= \det \Big[(J_+ A_1- \lambda I) (J_+ A_0- \lambda I) \\
                               & \hspace{1 cm}  - J_-^2 A_1 A_0 \Big] \nonumber \\
                              & := \det(G)
\end{align}
where to compactify the RHS we have defined
\begin{equation}
G := \begin{bmatrix}
    g_0   & g_1  & \dots & g_{N-1}\\
    g_{N-1} & g_0 &  \dots & g_{N-2} \\
    \hdotsfor{4} \\
    g_{1}  & g_2  & \dots & g_0
\end{bmatrix} \label{G}
\end{equation}
where
\begin{align}
& g_0 = \lambda^2 + \frac{J_+}{N} \lambda (N-1 - c_0) + \frac{J_-^2-J_+^2}{N} c_0 \\
& g_{(k>0)} = -\frac{J_+}{N} \lambda (1+c_k) + \frac{J_-^2-J_+^2}{N} c_k 
\end{align}
We want the determinent of $G$ which the product of its eigenvalues $\hat{\lambda}_j$ (not to be confused $M_{SPW}$'s eigenvalues $\lambda_i$ which we are trying to find),
\begin{align}
    \det(G) &= \prod_{j=0}^{N-1} \hat{\lambda}_j = 0
\end{align}
\noindent
Casting an eye back to Eq.~\eqref{G}, we see $G$ is a circulent matrix so its eigenvalues $\hat{\lambda}_j$ are known exactly! --
\begin{align}
\hat{\lambda}_j &= \sum_{k=0}^{N-1} g_k \beta_{j*k} \\
\hat{\lambda}_j &= \lambda^2 + \lambda J_+ - \frac{J_+}{N} \lambda \sum_{k=0}^{N-1} (1 + c_k) \beta_{j*k} \nonumber  \\
& \hspace{2 cm} + \frac{J_-^2 - J_+^2}{N} \sum_{k=0}^{N-1} c_k \beta_{j*k} \label{lam_hat}
\end{align}
where $\beta_k$ is the primitive root of unity as before. Note we mean the product $j*k$ in $\beta_{j*k} = \cos 2\pi j*k /N + \mathcal{I} \sin 2 \pi j*k /N$. Note also that while $G$ has $N$ eigenvalues $\hat{\lambda}$ (running from $j = 0, \dots N-1$), $M_{SPW}$ still has the required $2N$ eigenvalues $\lambda$ since each $\hat{\lambda_j}$ is quadratic in $\lambda$.

One last push remains: we simplify the summands in Eq.~\eqref{lam_hat} using some basic trig identities, find each $\hat{\lambda}_j$ for $j = 0, 1, \dots N-1$, and then set $\hat{\lambda}_j =0 $ (since $\det(G) = 0$ so each term in the product must be zero) which yields a quadratic equation for our target $\lambda_j$. 

We begin with $j = 0$ since it is distinguished from the other values $j$ takes. The first summand on the RHS of Eq.\eqref{lam_hat} becomes $\sum_{k=0}^{N-1} (1 + c_k) \beta_{0} = \sum_{k=0}^{N-1} (1 + c_k)(1) = N + 0 = N$. The second summand becomes $\sum_{k=0}^{N-1} c_k \beta_{j*k} = \sum_{k=0}^{N-1} c_k (1) = 0 $ (which follows from from Eq.~\eqref{cos_zero}). Plugging these in yields
\begin{align}
\hat{\lambda}_0 &= \lambda^2 + \lambda J_+ - \frac{J_+}{N} \lambda (N)  = \lambda^2 \label{lam_zero1}
\end{align}
Setting  $\hat{\lambda}_0 = 0$ gives our first eigenvalue
\begin{align}
\lambda_0 = 0 \label{lam_zero}
\end{align}
with multiplicity $2$.

Next we analyze $\hat{\lambda}_{j>0}$ all at once. First note
\begin{align}
    \sum_{k=0}^{N-1} \beta_{j*k} &= \sum_{k=0}^{N-1} \cos \frac{2 \pi j k}{N}  + \mathcal{I} \sin \frac{2 \pi j k}{N}  \\
    & = 0 + 0 \times \mathcal{I} = 0 \label{s4}
\end{align}
for all $j > 0$ (when $j = 0$ we get the simple sum $N-1$ which is why we considered the case $j=0$ separately). This implies the second summand in Eq.~\eqref{lam_hat} simplifies into the third summand $ \sum_{k=0}^{N-1} (1 + c_k) \beta_{j*k} = \sum_{k=0}^{N-1} c_k \beta_{j*k}$ when $j > 0$ which in turn becomes
\begin{align}
\sum_{k=0}^{N-1} c_k \beta_{j*k}  =\sum_{k=0}^{N-1} c_k c_{j*k} + \mathcal{I} c_k s_{j*k} \label{s5}
\end{align}
Using the standard formulas,
\begin{align}
\cos(a) \cos(b) &= \frac{1}{2} \Big( \cos(a+b) + \cos(a-b) \Big) \\
\cos(a) \sin(b) &= \frac{1}{2} \Big( \sin(a+b) - \sin(a-b) \Big)
\end{align}
we see 
\begin{align}
\sum_{k=0}^{N-1} c_{k} c_{j*k} &= \frac{1}{2} \sum_{k=0}^{N-1}  \Big( \cos \frac{ 2\pi(1+j)k}{N} +  \cos \frac{ 2\pi (1-j)k}{N} \Big) \label{y1} \\
\sum_{k=0}^{N-1} c_k s_{j*k} &= \frac{1}{2} \sum_{k=0}^{N-1} \Big( \sin \frac{ 2\pi(1+j)k}{N} +  \sin \frac{ 2\pi (1-j)k}{N} \Big)\label{y2}
\end{align}
where we have inverted the $c_j$ notation for clarity. Both the cosine terms in Eq.~\eqref{y1} and sine terms in Eq.~\eqref{y2} are zero (as per Eq.\eqref{s4}) except when the arguments are $0$. This occurs for $j = 1, -1$ (where $j = -1$ is interpreted modulo $N$ and equivalent to $j = N-1$. Since $\cos 0 = 1$ and $\sin = 0$, this yields 
\begin{align}
    \sum_{k=0}^{N-1} c_{k} c_{jk}  &= \frac{N}{2} (\delta_{j,1} +  \delta_{j, N-1}) \\
    \sum_{k=0}^{N-1} c_{k} s_{jk}  &= 0
\end{align}
where $\delta_{i,j} = 1 $ is the Kronecker delta. Applying the above to Eq.~\eqref{s5} gives
\begin{align}
\sum_{k=0}^{N-1} c_k \beta_{j*k}  = \frac{N}{2} (\delta_{j,1} + \delta_{j, N-1})
\end{align}
Plugging this into Eq.~\eqref{lam_hat} for $\hat{\lambda}_j$
\begin{align}
    & \hat{\lambda}_j = \lambda^2 + \lambda J_+ + \Big( - \frac{J_+}{N} \lambda  + \frac{J_-^2 - J_+^2}{N} \Big) (\delta_{j,1} + \delta_{j, N-1}) = 0 \label{lam_hat_1}
\end{align}
which holds for $j = 1, \dots, N-1$ since, recall, we have analyzed $j = 0$ separately and in that case found $\lambda = 0$ with multiplicity $2$ (Eq.~\eqref{lam_zero}). The desired $\lambda$ are the roots of the above equations. Both their values and multiplicities depend on $N$. For $N \geq 4$, a general pattern holds, but $N=2,3$ are special cases.

$N = 2$ is special because both of the Kronecker delta functions trigger at the same time: $ \delta_{j,N-1} = \delta_{j,2-1} = \delta_{j,1} = 1$. Eq.~\eqref{lam_hat_1} then becomes
\begin{align}
    & \hat{\lambda}_j = \lambda^2 + \lambda J_+ + 2 \Big( - \frac{J_+}{N} \lambda  + \frac{J_-^2 - J_+^2}{N} \Big) = 0 \\
    & \Rightarrow  \lambda = \pm \sqrt{J_+{-}^2 - J_-^2}
\end{align}
Table~\ref{lam_spw} row $1$ reports these eigenvalues along with the two zero eigenvalues $\lambda = 0$ which gives the required $2N = 4$ eigenvalues total.

$N = 3$ is special because precisely one of the two kronecker $\delta_{i,j}$ functions is on for all $j$; there is no $j$ for which both $\delta_{j,1}, \delta_{j,N-1}$ are zero simultaneously. For $j=1, \delta_{j,1} = 1$ and for $ j = N-1, \delta_{j,N-1} = \delta_{j,2} = 1$ (recall $j$ from from $1$ to $N-1$; so when $N = 2, j=1,2$ only). So we get the following for both $j=1,2$
\begin{align}
    & \hat{\lambda}_j = \lambda^2 + \lambda J_+ + \Big( - \frac{J_+}{N} \lambda  + \frac{J_-^2 - J_+^2}{N} \Big) = 0 \\
    & \Rightarrow  \lambda_{\pm} = \frac{1}{4} \left(-J_+ \pm \sqrt{9 J_+^2-8 J_-^2}\right)
\end{align}
and so the \textit{pair} $\lambda_{\pm}$ has multiplicity two as shown Table~\ref{lam_spw} row 2 (i.e. the full set is $\lambda_+, \lambda_+, \lambda_-, \lambda_-$). 

Finally, for $N \geq 4$, there are now intermediary values of $j = 2, \dots N-2$ for which $\delta_{j,1} = \delta_{j,N-1} = 0$ simultaneously in which case 
\begin{align}
    &\hat{\lambda}_j = \lambda( \lambda + J_+) = 0 \\
    &\Rightarrow \lambda = 0, -J_+
\end{align}
Table~\ref{lam_spw} row $3$ summarizes the $\lambda$'s and their multiplicities in this general case which finally completes our calculation. We have confirmed the expressions for $\lambda$ are correct by using Mathematica (notebook provided at \footnote{https://github.com/Khev/swarmalators/blob/master/1D/on-ring/regular/stability-static-phase-wave.nb}) to compute the eigenvalues of $M_{SPW}$ for $N = 2, \dots 6$ (for larger $N$, Mathematica starts to struggle).

\begin{table}[t]
 \begin{tabular}{| c | l |} 
 \hline
  $N$ & (Multiplicity, eigenvalue) \\ [0.5ex] 
 \hline
 $2$ & $ \lambda = (2,0), (1, \pm \sqrt{J_+^2 - J_-^2} )$ \\ 
 $3$ & $ \lambda = (2,0), (2, \left(-J_+ \pm \sqrt{9 J_+^2-8 J_-^2}\right) / 4)$ \\ 
 $4$ & $ \lambda = (3,0), (1,-J_+), (2,  \left(-J_+ \pm \sqrt{9 J_+^2-8 J_-^2}\right) / 4)$ \\ 
 $N\geq 4$ & $ \lambda = (0,N-1), (N-3,-J_+), (2, \left(-J_+ \pm \sqrt{9 J_+^2-8 J_-^2}\right) / 4)$ \\ 
 \hline
\end{tabular}
\caption{Spectrum of static phase wave for different population sizes. The cases $N = 2,3$ are distinguished. For $N \geq 4$ the pattern is fixed. Notice a multiplicity $1$ for eigenvalues that are complex conjugates $\lambda_{\pm}$ means one instance of the pair; a multiplicity of $2$ means two instances of the pair: $\lambda_+, \lambda_+, \lambda_-, \lambda_-$.  }
\label{lam_spw}
\end{table}

Now we use the expressions for $\lambda$ to predict the bifurcations of the static phase wave. For convenience we write out the expressions below (which are valid for $N > 2$. For $N = 2$, $\lambda = \sqrt{J_-^2 - J_+^2} > 0 $ for $K < 0$ meaning the static phase wave is unstable in this case. We will analyze the $N=2$ case fully in the next section). 
\begin{align}
    \lambda_0 &= 0 \\
    \lambda_1 &= -J_+ \\
    \lambda_2 &=  \frac{1}{4} \left(-J_+ \pm \sqrt{9 J_+^2-8 J_-^2}\right) \label{lam_hopf}
\end{align}
$\lambda_2$ exists for all $N \geq 3$ and triggers a Hopf bifurcation at $K_c = - J$ which comes from solving $Re(\lambda_2^{\pm}) = 0$ (the determinant $9J_+^2 - 8 J_-^2$ is negative here so $\lambda_2^{\pm}$ are complex conjugates). It also triggers a saddle node bifurcation at $K = 0$ which comes from solving $\lambda_2^+ = 0$ ($\lambda_2^-$ is never $0$). $\lambda_1$ exists for all $N \geq 4$ and triggers a saddle node bifurcation also at $K_c = -J$. These results imply the static phase wave is stable when
\begin{equation}
    -J < K < 0
\end{equation}
for all population sizes $N > 2$. To recap, at the left boundary $K=-J$, it destabilizes via a Hopf bifurcation for $N=3$ and a simultaneous Hopf and saddle node for $N \geq 4$. At the right boundary $K = 0$, it destabilizes via a saddle node for all $N \geq 3$.


\begin{figure}
    \centering
    \includegraphics[width= 0.75 \columnwidth]{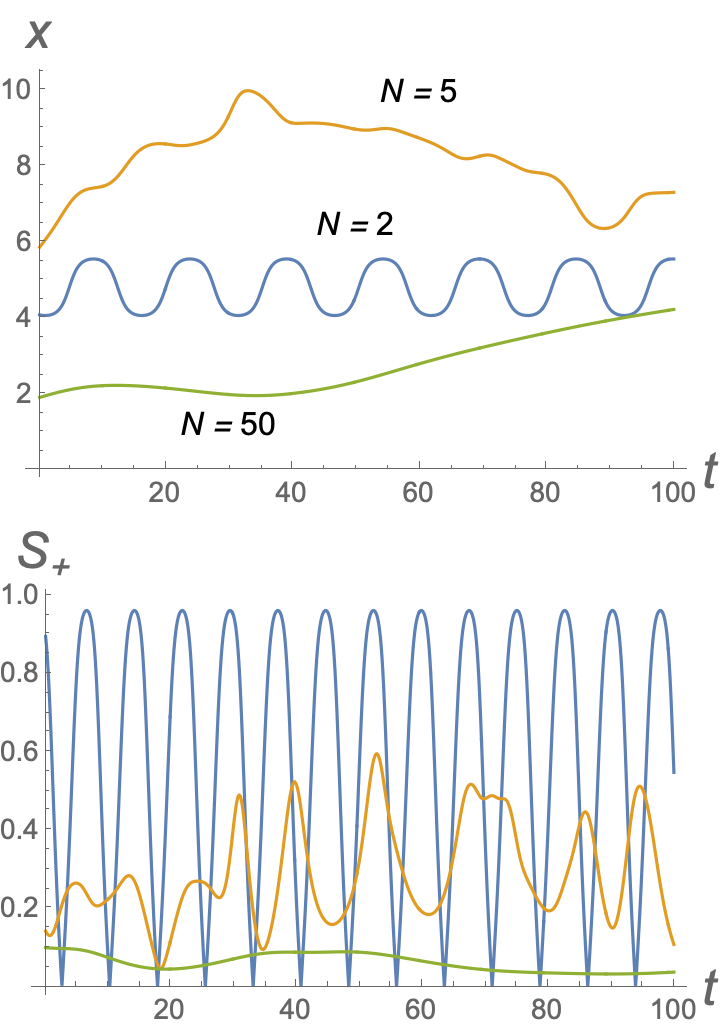}
    \caption{Dynamics in active async state. Top: Position $x(t)$ for a typical swarmalator for $N = 2, 5, 50$ swarmalators showing oscillatory behavior. For small $N$, the oscillations are smooth and fast. For larger $N$, they are irregular and slow. The dynamics of the swarmalator`s phase $\theta_i$ (unplotted) are similar. Bottom: time series of order parameter $S_{+}$. The behavior of $S_-$ is qualitatively the same. Simulation parameters were $(dt, T) = (0.1, 100)$ and $(J,K) = (1,-1.05)$.} 
    \label{active-phase-wave}
\end{figure}

\newpage
\textbf{Analysis of active async}.
The destabilization of the static phase wave via a Hopf bifurcation for $N \geq 2$ implies a limit cycle is born at $K < K_c$. Here swarmalators oscillate about their mean position in space and phase (Fig~\ref{states-1D}(d)) with little overall order $S_{\pm} \approx 0$ (Fig~\ref{order-parameters-1D-identical}) for most $N$. Figure~\ref{active-phase-wave} depicts the oscillations in $x(t)$ for a typical swarmalator for different population sizes. For small $N$, the oscillations are smooth and fast. For larger $N$, the oscillations are irregular and slow. (The oscillations in phase, unplotted, behave the same way). The bottom panel shows the dynamics of the rainbow order parameters $S_{\pm}$ are similar.

\begin{figure}
    \centering
    \includegraphics[width= 0.75 \columnwidth]{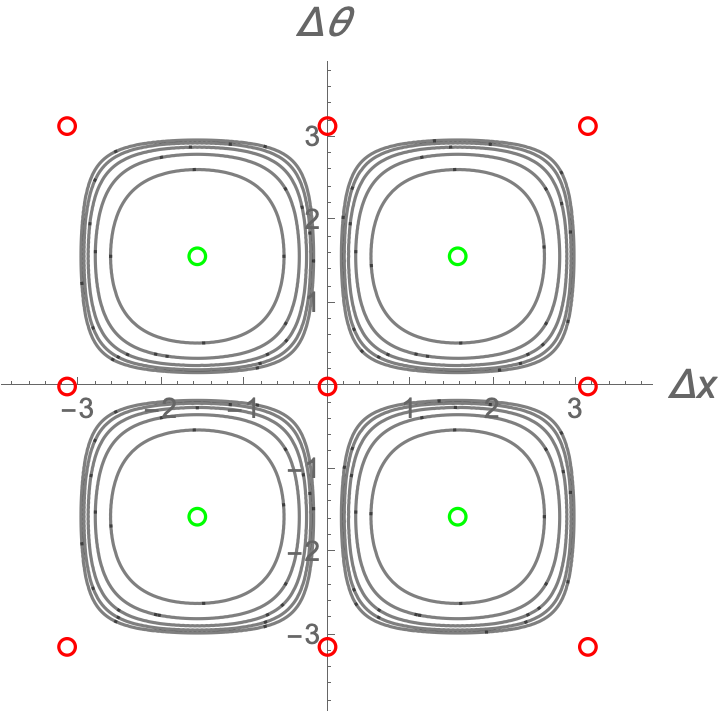}
    \caption{Phase space for $N = 2$ swarmalators in $(\Delta x, \Delta \theta)$ coordinates (Eqs.~\eqref{s1}, \eqref{s2}) for $(J,K) = (1,-1)$. Saddle point are shown as open red circles, nonlinear centers as open green circles. The family of periodic orbits, which correspond to the active async state, are given by Eq.~\eqref{s3} for $C = -6, \dots, 6$.} 
    \label{N_two}
\end{figure}

In the simple case $N = 2$ the active state can be studied using the standard transformation to mean and difference coordinates $(\langle y \rangle, \Delta y) = (y_1 + y_2, y_2 - y_1)$ where $y$ is a dummy variable. The mean positions and phases are invariant $\dot{\langle x \rangle} = \dot{\langle \theta \rangle} = 0$ because of the pairwise oddness of Eqs.~\eqref{eom-x},~\eqref{eom-theta} which implies $x_1 = - x_2$ and $\theta_1 = - \theta_2$. The differences, however, evolve according to
\begin{align}
    \dot{\Delta x} &= J \sin \Delta x \cos \Delta \theta \label{s1} \\
    \dot{\Delta \theta} &= K \sin \Delta \theta \cos \Delta x \label{s2}
\end{align}
These ODEs have 8 fixed points. For $K > 0$, the two fixed points $(\Delta x, \Delta \theta) = (0, \pi), (0,\pi)$ are stable nodes (which corresponds to static synchrony) while the rest are saddles. For $K < 0$, four nonlinear centers exist while the other fixed points are again saddles. The family of periodic solution surrounding the centers correspond to the active async state and can be found explicitly by dividing Eq.~\eqref{s1} by Eq.~\eqref{s2} and integrating,
\begin{align}
    &\frac{d{\Delta x}}{d{\Delta \theta}} = \frac{J \sin \Delta x \cos \Delta \theta}{K \sin \Delta \theta \cos \Delta x} \\
    & K \int \frac{\cos x}{ \sin x} dx = J \int \frac{\cos \theta}{ \sin \theta} d \theta \\
    & \sin x^K = C \sin \theta^J \label{s3}
\end{align}
For some constant $C$ determined by initial conditions. Figure~\ref{N_two} plots the contours of Eq.~\eqref{s3}, which correspond to the limit cycle of the active async state, along with the fixed point for $K = -1$. These are consistent with the oscillations in $x(t)$ in Figure~\ref{active-phase-wave}. They also tell us that the active async is stable for all $K < 0$ when $N=2$, which confirms our result from the last section that the static phase wave is unstable when $N=2$.

\begin{figure}
    \centering
    \includegraphics[width= 0.85 \columnwidth]{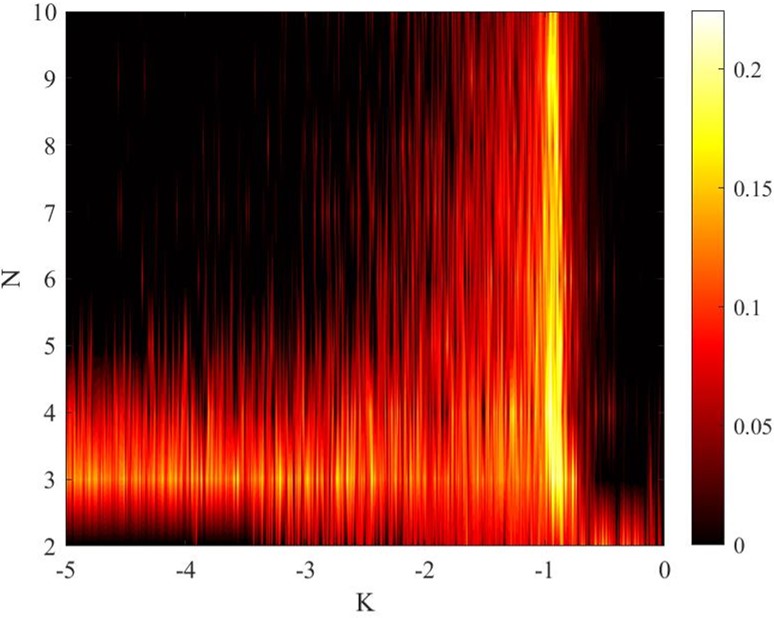}
    \caption{Mean velocity $\langle v \rangle = N^{-1} \sum_{i=1}^{N} v_i$. where $v_i = (v_x^2 + v_{\theta}^2)^{1/2}$ is the velocity of the $i$-th swarmalator, in the $(N,K)$ plane for $J = 1$ and $\nu_i = \omega_i = 0$. Simulation parameters were $(dt, T) = (0.1, 5000)$. The first $90\%$ of data were discarded as transients and the mean of the remaining $10\%$ are plotted.} 
    \label{v-N-K}
\end{figure}

For $N > 2$, analysis of the active async state becomes unwieldy so we numerically examine the state by plotting the mean velocity or activity $\langle v \rangle = N^{-1} \sum_{i=1}^{N} v_i$ where $v_i = (v_x^2 + v_{\theta}^2)^{1/2}$ is the velocity of the $i$-th swarmalator. Since all of the other collective states are stationary, $ \langle v \rangle $ is a natural order parameter for the active async state. Figure~\ref{v-N-K}, in which $J = 1$, shows the $\langle v \rangle $ persists for large $N$ when $K \approx K_c = -1$ (recall, we know the state must exist for \textit{some} $K$ for all finite $N$ via the Hopf bifurcation). For small coupling strengths $K \approx -2 $, however, the motion dies out for moderately large systems $N \gtrapprox 5$ and the final \textit{static async} state is born. 



\begin{figure}[h!]
    \centering
    \includegraphics[width= 0.85 \columnwidth]{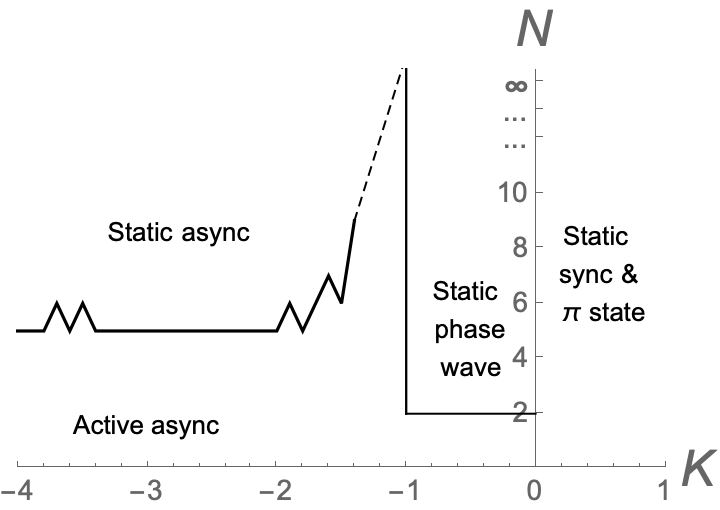}
    \caption{Pseudo-bifurcation diagram for model \eqref{eom-x}, \eqref{eom-theta} in $(K,N)$ space with $J = 1$ (pseudo because the population size $N$, a discrete quantitiy, is plotted on a continous axis; see main text). The $N \rightarrow \infty$ limit is represented by plotting $...$ on the $y$-axis. The dotted line schematically represents the bifurcation curve of the active async state and thus merges with the line $K = J = -1$ as $N \rightarrow \infty$.} 
    \label{bif-1D}
\end{figure}

\textbf{Analysis of static async}. Here the swarmalators sit at fixed points $(x_i^*, \theta_i^*)$ scattered uniformly in space and phase (Fig~\ref{states-1D}(e)) implying no global order $S_{\pm} = 0$ (Fig~\ref{order-parameters-1D-identical}). This state is hard to analyze for finite $N$ because for most population sizes $N \geq 3$, multiple configurations of fixed points $(x_i^*, \theta_i^*)$ exist for a given set of parameters $(J,K)$. Simply enumerating this family of fixed points is a hard problem -- it has not been done in the regular Kuramoto model -- never mind analyzing their stability. In the continuum limit $N \rightarrow \infty$, however, the stability may be analyzed since the state has a simple representation: $\rho(x,\theta,t) = 1/(4 \pi^2)$ where $\rho(x, \theta, t) dx d \theta$ gives the fraction of swarmalators with positions between $x + dx$ and phases between $\theta + d \theta$ at time $t$. The density obeys the continuity equation
\begin{equation}
    \dot{\rho} + \nabla . ( v \rho ) = 0
    \label{cont}
\end{equation}
And the velocity $v = (v_x, v_{\theta} )$ is given by the $N \rightarrow \infty$ limit of Eqs~\eqref{eom-x}, \eqref{eom-theta} 
\begin{align}
v_x &= J \int \sin(x' - x) \cos(\theta' - \theta) \rho(x',\theta',t) dx' d \theta' \\
v_{\theta} &= K \int \sin(\theta' - \theta) \cos(x' - x) \rho(x',\theta',t) dx' d \theta' \label{v}
\end{align}
The stability is analyzed by plugging a general perturbation
\begin{equation}
    \rho = \rho_0 + \epsilon \eta = (4 \pi^2)^{-1} + \epsilon \eta(x, \theta, t)
    \label{perturb}
\end{equation}
into the continuity equation and computing the spectrum. Since such analyses are standard, we put the details in Appendix C. The eventual result is
\begin{equation}
K < K_c = -J
\end{equation}
Recall this is only valid as $N \rightarrow \infty$. Interestingly, this result also proves the active async state disappears for $N \rightarrow \infty$ via a `squeeze' argument: since the static phase wave also loses stability at $K_c = -J$, there is no room left for the active async state to exist as $N \rightarrow \infty$.

\begin{figure}[h!]
    \centering
    \includegraphics[width= 0.85 \columnwidth]{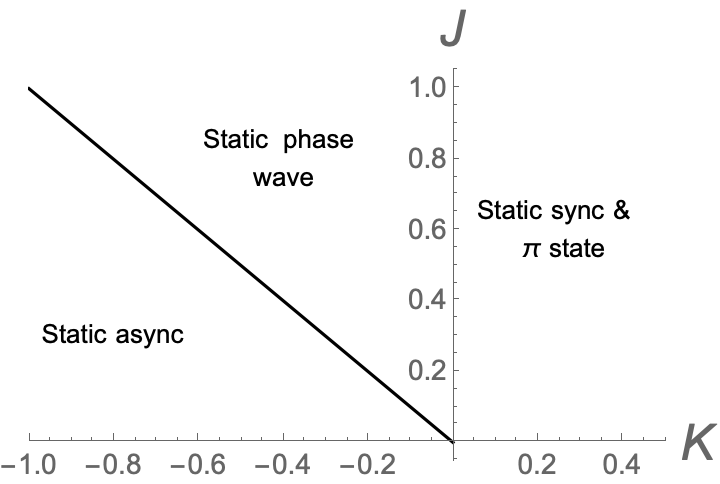}
    \caption{Bifurcation diagram in $(J,K)$ space as $N \rightarrow \infty$.} 
    \label{bif-1D-infinite-N}
\end{figure}

This completes our analysis. Figure~\ref{bif-1D} summarizes our findings in a pseudo bifurcation diagram in $(K,N)$ space for $J = 1$(pseudo because $N$ is a discrete quantity and so should plot by plotted on a continuous axis, but we wanted to show the active async state dependence on $N$). The boundary of the active async state for $N \geq 3$ is found numerically by finding the first time $\langle v \rangle = 0$ which indicates the swarmalators are no longer active. Figure~\ref{bif-1D-infinite-N} reports the bifurcation diagram in $(J,K)$ space when $N \rightarrow \infty$.

\section{Match to real world swarmalators}
Swarmalators are defined as entities with a two-way interaction between swarming and synchronization \cite{o2017oscillators}. Below we list examples which (i) appear to meet this definition \footnote{We say `appear' to meet these definition because what is means to prove a bidirectional space-phase coupling in an experimental system is somewhat ambiguous. Most experimental studies of swarmalators either infer a space-phase coupling exists based on the observations (like the microswimmers \cite{quillen2021metachronal,yang2008cooperation,yuan2014gait} we describe; \cite{yuan2014gait} do a particularly comprehensive study), but do not specify if the coupling bidirectional. Other studies demonstrate a two-way coupling in the sense that they encoding it in a model that produces behavior similar to observed data (like magnetic domain walls \cite{hrabec2018velocity} or myxobacteria \cite{igoshin2001pattern}). In other words, we're trying to be conservative in any claims we make.}, (ii) swarm in ring-like geometries, and (iii) display collective behavior similar to the ring model.

\textbf{Sperm} are a classic microswimmer which swarm in solution and sync their tail gaits \cite{yang2008cooperation}. Sperm collected from ram semen and confined to 1D rings bifurcate from an isotropic state (analogues to the static async state) to a vortex state in which sperm rotate either clockwise or anticlockwise \cite{creppy2016symmetry} which implies their positions and orientations are splayed just like the static phase wave (recall the static phase wave is really a state of uniform rotation and only static in the frame co-moving with the natural frequencies $\omega, \nu $) \cite{creppy2016symmetry}. Moreover, the transition has associated transient decay of rotation velocity (see Fig 4a in \cite{creppy2016symmetry}) consistent with a Hopf bifurcation, just like the ring model \eqref{lam_hopf}. Note unlike other studies of syncing sperm \cite{yang2008cooperation}, here the phase variable is the sperm's orientation, not its tail rhythm.

\textbf{Vinegar eels} are a type of nematode found in beer mats and the slime from tree wounds \cite{quillen2021metachronal,quillen2021synchronized}. They can be considered swarmalators because they sync the wriggling of their heads, swarm in solution, and it seems likely based on their behavior said sync and swarming interact \cite{quillen2021metachronal,quillen2021synchronized} (neighbouring eels sync more easily than distant eels, so sync interacts with swarming, and sync'd eels presumable affect each local hydrodynamic environment and thereby affect each other's movements, so swarming interacts with sync). When confined to 2D disks, they seek out the 1D ring boundary forming metachronal waves in which the phase of their gait and their spatial positions around the ring are splayed similar to the static phase wave \cite{quillen2021metachronal,quillen2021synchronized} (although note the winding number for the metachronal waves is $k>1$; a full rotation in physical space $x$ produces $k > 1$ rotations in phase $\theta$).

\textbf{C. elegans} are another type of microswimmer which also swarm and sync the gait of their tails. When confined to 1D channels they form synchronous clusters analogous to the static sync state \cite{yuan2014gait}. Though not strictly consistent with the ring model, a channel being a 1D line as opposed to a 1D ring, we mention them here because it is natural to expect sync clusters would persist in a 1D ring too.

\section{Discussion}
Aim one of the paper was to model real-world swarmalators swarming in 1D. This was a success. The model captured the behavior of vinegar eels, sperm, and C. elegans. Even so, the model failed to capture the phenomenology of other 1D swarmalators such as the the two-cluster states of bordertaxic Japanese tree frogs \cite{aihara2014spatio} or the cluster dynamics of synthetic microswimmers \cite{guzman2016fission}. Crafting a model that mimics these systems is a challenge for future research.

Aim two was to provide a stepping stone -- i.e. the ring model --  to an analytic understanding of the 2D swarmalator model's phenomenology, in particular its bifurcations. To this end, we studied the destabilization of the static async state,  deriving a 1D version of $K_1$. But an analogue of the active phase wave state did not appear in the ring model \footnote{We note the splintered phase wave and active phase waves are realized in a 1D ring model presented in the Supplementary Information of \cite{o2017oscillators}. But that model does not have the clean format of the current model Eqs.\eqref{eom_xi},\eqref{eom_eta} which, being so simple, seem the natural model to study 1D swarmalator phenomena. Moreover, the analogues of the splintered phase wave and active phase wave states there are unsteady, the order parameters varying in time. So in that sense that are not simpler warm-up versions of the states in 2D.}, so we could not study the second bifurcation at $K_2$ beyond which $S_+$ declines (Figure~\ref{order-parameters-2D}). On the upside, a 1D analogue of the static phase wave was observed, and were able to specify its stability (in the 2D model this state was only observed for $K=0$ and its stability was not calculated). Moreover, this stability calculation certified the existence of the new active async state (we say certify because we originally thought the state was just a long transient or perhaps a numerical artefact; our results prove it exists for all finite $N$) in which the swarmalators execute noisy, Brownian-like motion. What's interesting here is that this motion occurs for identical, noise-free swarmalators. This suggest any jittery behavior observed in real-world swarmalator systems may arise purely from the cross-talk between units' tendency to sync and swarm, and not from thermal agitation or other forms of noise. 

Taken together, our analytic findings take us one step closer to understanding the bifurcations of the swarmalator model which for now remain unexplained.


We would love to see future work analyze the ring model using OA theory \cite{ott2008low,pikovsky2015dynamics,engelbrecht2020ott,chen2017hyperbolic}. A breakthrough from \underline{O}tt and \underline{A}ntonsen, the theory states that the density of the infinite-$N$ Kuramoto model $\rho(\theta, t)$ has an invariant manifold of poisson kernels which allows dynamics for the classic sync order parameter $Z := N^{-1} \sum_j e^{i \theta_j}$ to be derived explicitly. This is a big win. It reduces an $N >> 1$-dimensional nonlinear system to a simple 2D system (one ODE for the complex quantity $Z$) -- a drastic simplification which effectively solves the Kuramoto model. Given the ring model in $(\xi, \eta)$ coordinates (Eqs.\eqref{eom_xi},\eqref{eom_eta}) resembles the Kuramoto model so closely, we suspect it may be solved with OA theory's magic: If regular oscillators are defined on the unit circle $\theta_i \in S_1$ and have an invariant manifold of poisson kernels, could ring swarmalators, defined on the unit torus $(x, \theta) \in (S^1, S^1)$, have an invariant manifold of some `toroidal' poisson kernel? If so, explicit dynamics for the rainbow order parameters $W_{\pm}$ may be derivable and in that sense the ring model solved exactly.

\bibliographystyle{apsrev}


\appendix

\section{Basin of attractions for static sync and $\pi$-state}
\begin{figure}
    \centering
    \includegraphics[width=\columnwidth]{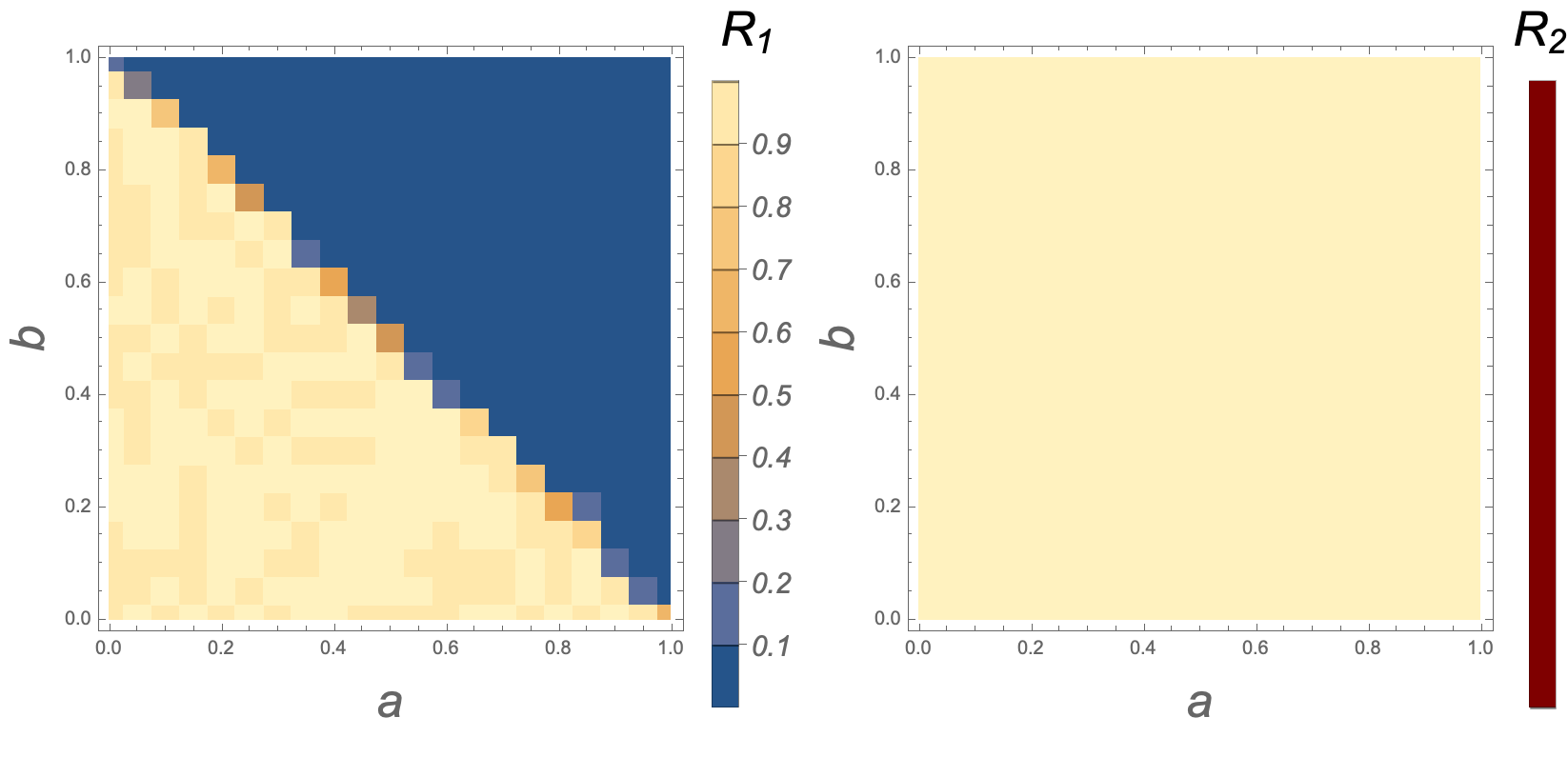}
    \caption{Left panel: $R_1(a,b)$. Right panel: $R_2(a,b)$. Data were collected by integrating Eqs.\eqref{eom-x}, \eqref{eom-theta} using an RK4 method with $(dt, T) = (0.1, 500)$ for $N = 100$ swarmalators.}
    \label{basins-of-attraction}
\end{figure}

Are the static sync and $\pi$ states the only stable collective states when $K > 0$? Answering this question conclusively is difficult since it would mean integrating the equations of motion for every initial condition in the phase space $T^N = S^N \times S^N$. Figure~\ref{basins-of-attraction} explores a subset of $T^N$: initial position $x_i$ spaced uniformly on $(0, a \pi)$ and initial phases $\theta_i$ spaced uniformly on $(0, b \pi)$ where $0 \leq a,b \leq 2$ and then perturbed slightly with noise of order $10^{-3}$ to both $x_i, \theta_i$. When $(a,b) = (2,2)$ these initial conditions correspond to swarmalators spread out evenly over the unit circle in both space and phase. When $a,b < 2$, the swarmalators are spread out over subsets of the unit circle. 

We use the Daido order parameters $R_1 e^{i \phi_1} := N^{-1} \sum_j e^{i \theta_j} $ and $R_2 e^{i \phi_2} = N^{-1} \sum_j e^{2 i \theta_j}$ to distinguish between the static and $\pi$ states. The conditions for each state are
\begin{enumerate}
    \item \text{Condition for static sync:} $(R_1, R_2) = (1,1)$
    \item \text{Condition for $\pi$-state:} $(R_1, R_2) = (0, 1)$
\end{enumerate}
While $R_2 = 1$ for both the static sync and $\pi$ states, and thus can't distinguish between them, we include it to rule out the existence of some other collective state for which $R_2 \neq 0$. 

Figure~\ref{basins-of-attraction} shows that the static sync is stable when $(0 \leq a \leq 1) \cap (0 \leq b \leq 1)$ and the $\pi$-state is stable in the remaining space. No other collective states were observed.


\section{Connection of ring model to 2D swarmalator model}

\begin{figure}
    \centering
    \includegraphics[width=\columnwidth]{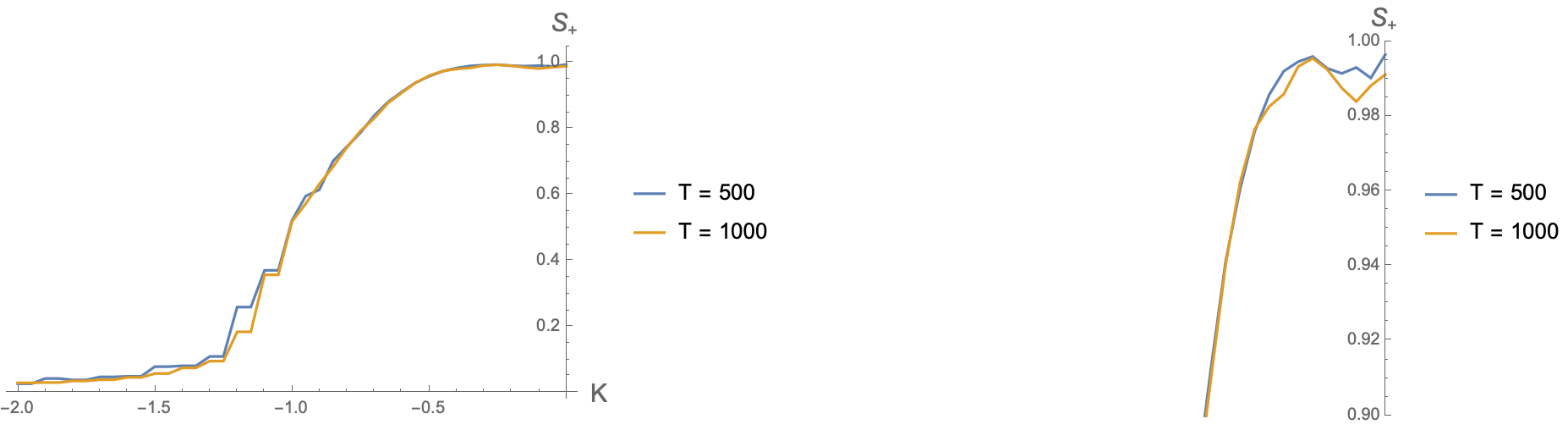}
    \caption{Left panel: $S_+$ for the unit vector model which has the same qualitative shape as that of the linear parabolic model presented in Figure~\ref{order-parameters-2D}. Results for $T = 500, 1000$ are shown to establish convergence. Right panel: zoom in of $S_+(K)$ showing the non-monotonic dip. Results were generated using an RK4 solver with $dt = 0.1$.}
    \label{S-linear-parabolic}
\end{figure}

\begin{widetext}
Here we show how the ring model is contained within the 2D swarmalator model which is given by
\begin{align}
&\dot{\mathbf{x}}_i = \mathbf{v}_i + \frac{1}{N} \sum_{j=1}^N \Big[ \mathbf{I}_{\mathrm{att}}(\mathbf{x}_j - \mathbf{x}_i)F(\theta_j - \theta_i)  - \mathbf{I}_{\mathrm{rep}} (\mathbf{x}_j - \mathbf{x}_i) \Big],   \\ 
& \dot{\theta_i} = \omega_i +  \frac{K}{N} \sum_{j=1}^N H_{\mathrm{\mathrm{att}}}(\theta_j - \theta_i) G_{\sigma}(\mathbf{x}_j - \mathbf{x}_i)
\end{align}
In \cite{o2017oscillators}, the choices $I_{att} = x / |x|$, $I_{rep} = x / |x|^2$, $F(\theta) = 1 + J \cos(\theta)$, $G(x) = 1 / |x|$, $H_{att}(\theta) = \sin(\theta)$ were made. However, choosing linear spatial attraction $I_{att}(x) = x$, inverse square spatial repulsion $I_{rep}(x) = x / |x|^2$ and truncated parabolic space-phase coupling $G(x) = (1 - |x|^2/ \sigma^2) H_{heaviside}({\sigma - |x|})$ 
\begin{align}
&\dot{\mathbf{x}}_i = \frac{1}{N} \sum_{ j \neq i}^N \Bigg[ \mathbf{x}_j - \mathbf{x}_i \Big( 1 + J \cos(\theta_j - \theta_i)  \Big) -   \frac{\mathbf{x}_j - \mathbf{x}_i}{ | \mathbf{x}_j - \mathbf{x}_i|^2}\Bigg] \label{linear_parabolic1} \\ 
& \dot{\theta_i} = \frac{K}{N} \sum_{j \neq i}^N \sin(\theta_j - \theta_i ) \Big(1 - \frac{| \mathbf{x}_j - \mathbf{x}_i|^2}{\sigma^2} \Big) H_{heaviside}(\sigma - |\mathbf{x}_j - \mathbf{x}_i|) \label{linear_parabolic2}
\end{align}
\noindent
gives the same qualitative behavior but is nicer to work with analytically. 

First we show it demonstrates the same behavior. Figure~\ref{states-2D} shows its collective states are the same as those of the original model, and Figure~\ref{order-parameters-2D} shows its order parameters $S_{\pm}(K)$ have the same shape: monotonic increase from $(K_1, K_2)$ as $K \rightarrow 0^{-1}$, monontonic decrease from $(K_2, 0)$ as $K \rightarrow 0^{-1}$, and then a discontinuous jump to $S_+ = 1$ at $K=0$ ($K < 0$ is a singular perturbation). Note the plot of $S_{\pm}$(K) in Figure 6 of \cite{o2017oscillators}, the monotonic decrease on $(K_2,0)$ of $S_+$ is slight and hard to see, so we plot in for finer $K$ and show a zoom in for small $K$ in Figure~\ref{order-parameters-2D-unit-vector}. 

Now we show the `linear parabolic` model, so called because $I_{att} = x$ and $G(x)$ is a parabolic, is cleaner analytically. In polar coordinates it takes form

\begin{align*}
\dot{r_i}&=  H_r(r_i, \phi_i) - J r_i R_0 \cos \Big( \Psi_0 - \theta_i \Big) + \frac{J}{2} \Bigg[ \tilde{S}_+ \cos \Big( \Phi_+- (\phi_i+\theta_i) \Big) 
 +  \tilde{S}_{-}\cos \Big( \Phi_{-} - (\phi_i-\theta_i) \Big) \Bigg] \\
\dot{\phi_i}&=  H_{\phi}(r_i, \phi_i)  + \frac{J}{2 r_i} \Bigg[ \tilde{S}_+ \sin \Big( \Psi_+ - (\phi_i+\theta_i) \Big)  + \tilde{S}_- \sin \Big( \Psi_{-} -  (\phi_i- \theta_i) \Big) \Bigg] \\
\dot{\theta_i} &=  K \Big(1- \frac{r_i^2}{\sigma^2} \Big) R_0 \sin (\Phi_0 - \theta_i) - \frac{K}{\sigma^2} R_1 \sin(\Phi_1 - \theta_i) + \frac{K r_i}{\sigma^2} \Bigg[ \tilde{S}_+ \sin \Big( \Psi_+ - (\phi_i+\theta_i) \Big) - \tilde{S}_- \sin \Big( \Psi_- - (\phi_i- \theta_i) \Big) \Bigg]
\end{align*}
\noindent
where
\begin{align}
H_r(r_i, \phi_i) &=  \frac{1}{N} \sum_{j} \Big( r_j \cos(\phi_j - \phi_i) - r_i \Big) ( 1 - d_{ij}^{-2} )  \label{Hr} \\
H_{\phi}(r_i, \phi_i) &=  \frac{1}{N} \sum_{j} \frac{r_j}{r_i} \sin(\phi_j - \phi_i) ( 1 - d_{ij}^{-2} ), \label{Hphi}  \\
Z_0 = R_0 e^{i \Psi_0}&=  \frac{1}{N} \sum_{j}  e^{i \theta_j}, \label{Qzero} \\
\hat{Z}_0 = \hat{R}_0 e^{i \hat{\Psi}_0}&=  \frac{1}{N} \sum_{j \in N_i}  e^{i \theta_j}, \label{Qzero} \\
Z_2 = R_2 e^{i \Psi_2}&=  \frac{1}{N} \sum_{j} r_j^2 e^{i \theta_j}, \label{Q1} \\
\hat{Z}_2 = \hat{R}_2 e^{i \hat{\Psi}_2}&=  \frac{1}{N} \sum_{j \in N_i} r_j^2 e^{i \theta_j}, \label{Q2} \\
\tilde{W}_{\pm} = \tilde{S}_{\pm} e^{i \Psi_{\pm}}&=  \frac{1}{N} \sum_{j} r_j e^{i (\phi_j \pm \theta_j)} 
\label{Qplusminus} \\
\hat{W}_{\pm} = \hat{S}_{\pm} e^{i \hat{\Psi}_{\pm}}&=  \frac{1}{N} \sum_{j \in N_i} r_j e^{i (\phi_j \pm \theta_j)}
\end{align}
\noindent
where the $\hat{Z_0}, \dots$ order parameters are summed over all the neighbours $N_i$ of the $i$-th swarmalator: those within a distance $\sigma$. Notice that rainbow order parameters $\tilde{W}$ here are weighted by the radial distance $r_j$, which is not the case for the ring model Eqs.\eqref{x_eom_model}, \eqref{theta_eom_model} (that's why we put a tilde over the W).  Assuming $\sigma > max(d_{ij})$, we can set $\hat{Z_0} = Z_0, \hat{Z}_1 = Z_1, \hat{W_\pm} = W_{\pm}$. Then $S_{\pm} \sin(\Phi_{\pm} - (\phi \pm \theta) )$ etc of the ring model starting to emerge. If we assume there is no global synchrony $Z_0 = Z_2 = 0$, which happens generically in the frustrated parameter regime $K < 0, J > 0$, and transform to $\xi_i = \phi_i + \theta_i$ and $\eta_i = \phi_i - \theta_i$ coordinates the ring model is revealed (the terms in the square parentheses in the latter two equations.)
\begin{align}
\dot{r_i}&= \tilde{\nu}(r_i) +  \frac{J}{2} \Bigg[ \tilde{S}_+ \cos \Big( \Phi_+- \xi_i \Big) + \tilde{S}_{-}\cos \Big( \Phi_{-} - \eta_i \Big) \Bigg] \\
\dot{\xi_i}&= \tilde{\omega}(r_i, \phi_i) + \Bigg[ J_+(r_i) \tilde{S}_+ \sin \Big( \Psi_+ - \xi_i \Big)  + J_- (r_i) \tilde{S}_- \sin \Big( \Psi_{-} -  \eta_i \Big) \Bigg]   \\
\dot{\eta_i} &= \tilde{\omega}(r_i, \phi_i) +  \Bigg[ J_-(r_i) \tilde{S}_+ \sin \Big( \Psi_+ - \xi_i \Big) - J_+(r_i) \tilde{S}_- \sin \Big( \Psi_- - \eta_i \Big) \Bigg]
\end{align}
where
\begin{align}
\tilde{\nu}(r_i, \phi_i) &= H_r(r_i, \phi_i) \\
\tilde{\omega}(r_i, \phi_i) &= H_{\phi}(r_i, \phi_i) \\
J_{\pm}(r_i) &= \frac{J}{2 r_i} \pm \frac{K r_i}{\sigma^2}
\end{align}
\end{widetext}


\newpage
\section{Stability of static async state}

The density obeys the continuity equation
\begin{equation}
    \dot{\rho} + \nabla . ( v \rho ) = 0
    \label{cont}
\end{equation}
And the velocity $v = (v_x, v_{\theta} )$ is given by the $N \rightarrow \infty$ limit of Eqs~\eqref{eom-x}, \eqref{eom-theta} 
\begin{align}
v_x &= J \int \sin(x' - x) \cos(\theta' - \theta) \rho(x',\theta',t) dx' d \theta' \\
v_{\theta} &= K \int \sin(\theta' - \theta) \cos(x' - x) \rho(x',\theta',t) dx' d \theta' \label{v}
\end{align}
Consider a perturbation around the static async state $\rho_0 = 1 / (4 \pi^2)$
\begin{equation}
    \rho = \rho_0 + \epsilon \eta = (4 \pi^2)^{-1} + \epsilon \eta(x, \theta, t)
    \label{perturb}
\end{equation}
Normalization requires $\int \rho(x,\theta) = 1$ which implies 
\begin{equation}
    \int \eta(x,\theta,t) dx d \theta = 0
    \label{norm}
\end{equation}
The density ansatz \eqref{perturb} decomposes the velocity
\begin{equation}
    v = v_0 + \epsilon v_1 =  \epsilon v_1
    \label{vel}
\end{equation}
\noindent
where $v_0$ is velocity in the static async state $v_{0} = 0$. The perturbed velocity $v_1$ is given by Eqs~\eqref{v} with $\rho$ replaced by $\eta$. Plugging Eqs~\eqref{perturb}, \eqref{vel} into \eqref{cont} yields
\begin{equation}
    \dot{\eta} + \rho_0 (\nabla . v_1) = 0
    \label{cont1}
\end{equation}
To tackle this, first write the $v_1$ in terms of the order parameters $W_{\pm}$
\begin{align}
v_x^1 &= \frac{J}{2} \text{Im} \Big(  W_+^1 e^{-i(x+\theta)} + W_-^1 e^{-i (x -\theta)} \Big) \\
v_{\theta}^1 &= \frac{K}{2} \text{Im} \Big(  W_+^1 e^{-i(x+\theta)} - W_-^1 e^{-i (x -\theta)}  \Big)  \label{v}
\end{align}
\noindent
where the perturbed order parameters are 
\begin{equation}
W_{\pm}^1 = \int e^{i (x' \pm \theta')} \eta(x',\theta') dx' d \theta'    
\end{equation}
The divergence term is $\nabla . v_1. = \partial_x v_x + \partial_{\theta} v_{\theta}$ is easy to compute since only exponential terms
\begin{equation}
    \nabla . v_1 = - \frac{J+K}{2} \text{Re} \Big(  W_+^1 e^{-i(x+\theta)} + W_-^1 e^{-i (x -\theta)}  \Big)
\end{equation}
Plugging this into the evolution equation for $\eta$ Eq.~\eqref{cont1} yields 
\begin{equation}
    \dot{\eta}(x,\theta,t) = \frac{J+K}{8 \pi^2} \text{Re} \Big(  W_+^1 e^{-i(x+\theta)} + W_-^1 e^{-i (x -\theta)}  \Big)
\end{equation}
Using $\text{Re} \; z = \frac{1}{2}( z + \bar{z} )$ we get
\begin{align}
    \dot{\eta} &= \frac{J+K}{16 \pi^2}  \Big(  W_+^1 e^{-i(x+\theta)} + \bar{W}_+^1 e^{i(x+\theta)} + W_-^1 e^{-i (x -\theta)} \nonumber \\ 
    & + \bar{W}_-^1 e^{i (x -\theta)} \Big) \label{eta-dot}
\end{align}
Only a few Fourier modes are distinguished so we expand $\eta(x,\theta,t)$ in terms of complex exponentials
\begin{align}
    \eta &= (2 \pi)^{-1}  \Big( \alpha_{0,0}(t) + \alpha_{0,1}(t)e^{i \theta} + 
    \alpha_{1,0}(t) e^{i x} \\
    +& \sum_{n=1}^{\infty} \sum_{m=1}^{\infty} \bar{\alpha}_{n,m}(t) e^{i (n x + m \theta)} + \bar{\beta}_{n,m}(t) e^{i (n x- m \theta)} \nonumber \\ 
    & + c.c. \Big)
\end{align}
where $c.c$ denotes the complex conjugate. Unconventionally, we associate the complex conjugate $\bar{\alpha}, \bar{\beta}$ with the first harmonic. This is so that the order parameters are expressed in terms of $W_+^1 = \alpha_{1,1}(t)$ and $W_-^1  = \beta_{1,1}(t)$ (i.e. without the bar overhead). Also, notice the normalization condition~\eqref{norm} implies $\alpha_{0,0}(t) = 0$.

Integrating Eq.~\eqref{eta-dot} with respect to $\int (.) e^{i(n x \pm n \theta)}$ extracts the evolution equations for each mode. The ones of interest are
\begin{align}
    \dot{W}_{\pm}^1 = \frac{J+K}{4 \pi} W_{\pm}^1
\end{align}
and $\dot{\alpha}_{n,m} = \dot{\beta}_{n,m} = 0$ for all $(n,m) \neq (1,1)$. Setting $W_{\pm}^1 = w_{\pm}^1 e^{\lambda_{\pm} t}$ yields
\begin{align}
    \lambda_{\pm} = \frac{J+K}{4 \pi}
\end{align}
which implies the static async state is stable for 
\begin{align}
K < K_c = -J
\end{align}
consistent with simulations.

\end{document}


\title{Supplemental materials}

\maketitle

\section{General perturbation}
The continuity equation in one spatial dimension is
\begin{align}
    \dot{\rho} = - \partial_{x}( v \rho)  \\
    \dot{\rho} = - v \frac{\partial \rho}{\partial x} - \rho \frac{\partial v}{\partial x}.
\end{align}
Consider a general perturbation
\begin{align}
    \rho = \rho_0(x) + \epsilon \eta(x,t) \\
    v = v_0(x) + \epsilon v_1(x,t)
\end{align}
Subbing the perturbation into the continuity equations gives
\begin{align}
    \epsilon \dot{\eta} =  - (v_0 + \epsilon v_1)( \frac{\partial \rho_0}{\partial x} + \epsilon \frac{\partial \eta}{\partial x} ) - (\rho_0 + \epsilon \eta)( \frac{\partial v_0}{\partial x} + \frac{\partial v_1}{\partial x})
\end{align}
Collecting at $O(1)$ and $O(\epsilon)$
\begin{align}
   0 &=   v_0 \frac{\partial \rho_0}{\partial x}  + \rho_0 \frac{\partial v_0}{\partial x} \label{e1}  \\
  \dot{\eta} &=  - \Big( v_1 \frac{\partial \rho_0}{\partial x} + v_0 \frac{\partial \eta}{\partial x} \Big) - \Big( \eta \frac{\partial v_0}{\partial x} + \rho_0 \frac{\partial v_1}{\partial x} \Big) \label{e2}
\end{align}
Eq.~\eqref{e1} asserts $v_0$ and $\rho_0$ must be an equilibrium state we are perturbing around. Eq.~\eqref{e2} has all the information about the stability.

\section{Warm up 1}
Now we consider the system
\begin{align}
    \dot{x_i} = -x_i \label{warm_up_1}
\end{align}
Which has fixed points $x_i = 0$ with $\lambda = -1$. At the density level, $\rho_0(x) = \delta(x)$. We see to study the stability of $\rho_0$. I'm unsure how to specify $v_0$ etc. My belief is that $v_0 = 0$ (since no particles move in the fixed point) and $v_1 = -x$.

\subsection{Guess 1}
\begin{align}
   v_0 &= 0 \\
   v_1 &= -x \\
   \rho_0 &= \delta(x) 
\end{align}
Now we sub these into Eqs~\eqref{e1} and \eqref{e2}. Eq.~\eqref{e1} gives $0 = 0$. Eq.~\eqref{e2} gives
\begin{align}
    \dot{\eta} &= - \Big( -x \delta'(x) \Big) - \Big( \delta(x) (-1) \Big) \\
    \dot{eta} &= x \delta'(x) + \delta(x)
\end{align}
Recalling that $x \delta'(x) = -\delta(x)$ (integration by parts) we get
\begin{align}
    \dot{\eta} &= 0
\end{align}
The perturbation is neutrally stable. I was expecting $\lambda = -1$, but maybe $\lambda = 0$ is genuine? Or maybe setting $v_0 = 0$ is wrong. I'll set $v_0 = -x$ and see what happens, although this seems weird: there are no particles at $x \neq$ so there should be no velocity there. I guess I need to think about the Eulerian interpretation of $v(x) = -x$; if the density has compact support, does $v(x)$ too? Anyway, I'll see what happens.

\subsection{Guess 2}
\begin{align}
   v_0 &= -x \\
   v_1 &= -x \\
   \rho_0 &= \delta(x) 
\end{align}
Now we sub these into Eqs~\eqref{e1} and \eqref{e2}. Eq.~\eqref{e1} gives $0 = 0$. Eq.~\eqref{e2} gives
\begin{align}
    \dot{\eta} &= - \Big( -x \delta'(x) - x \frac{\partial \eta}{ \partial x} \Big) - \Big( \delta(x) (-1) - \eta \Big) \\
    \dot{\eta} &= \eta + x \frac{\partial eta}{ \partial x}
\end{align}
This looks wrong. It means $\eta$ grows for all $x$? Can solve the PDE for $\eta(x,t)$ exactly. For $\eta(x,0) = \pm 1$ we get $\eta(x,t) = \pm e^t$. So, setting $v_0 = -x$ must be wrong. So I conclude that the stability of Eq.\eqref{warm_up_1} is neutral...

\section{Warm up 2}
Now we consider the linearized Kuramoto model
\begin{align}
    \dot{x_i} = \frac{K}{N} \sum_j (x_j -x_i) \label{warm_up_2}
\end{align}
Which has fixed points $x_i = C = 0$ WLOG with $\lambda = -1$. At the density level, $\rho_0(x) = \delta(x)$. Now, notice $\langle x_i \rangle = 0$, which implies $\rho(x)$ and therefore $\eta(x)$ must be even: $\int x \eta(x) d x = 0$. This simplifies 
\begin{align}
    \dot{x_i} = -K x_i \label{warm_up_2}
\end{align}
Which is identical to warm up one Eq~\eqref{warm_up_1}. So nothing changes...? So we are ready for the locked state of the Kuramoto model.

\section{Locked state of kuramoto model with identical oscillators}
Now we consider the linearized Kuramoto model
\begin{align}
    \dot{\theta_i} &= \frac{K}{N} \sum_j \sin( \theta_j -\theta_i) \label{eom} \\
    v(\theta) &= K \int \sin(\theta' - \theta) d \theta'
\end{align}
Which has fixed points $\theta_i = C $ WLOG with $\lambda = -K$ and at the density level, $\rho_0(\theta) = \delta(\theta-C)$. As before, I'm guessing
\begin{align}
   v_0 &= 0 \\
   v_1 &= K \int \sin(\theta' - \theta) \eta(\theta') d \theta' \\
   \rho_0 &= \delta(\theta) 
\end{align}
Subbing this into Eq.~\eqref{e2} gives
\begin{align}
    \dot{\eta} &= - \Big(v_1 \delta'(\theta-C) \Big) - \Big( \delta(x-C) \frac{\partial v_1}{\partial \theta} \Big) \\
    \dot{\eta} &= - K \delta'(\theta-C) \int \sin(\theta' - \theta) \eta(\theta') d \theta'  + K \delta(\theta-C) \int \cos(\theta' - \theta) \eta(\theta') d \theta' 
\end{align}
Notice the sign change in the second term since the derivative on $v_1$ is with respect to $\theta$ and not $\theta'$. Defining $r_1 e^{i \psi_i} = \int e^{i \theta'} \eta(\theta',t) d \theta'$ we get
\begin{align}
    \dot{\eta} &= - K \delta'(\theta-C) r_1 \sin(\psi_1 - \theta)  + K  \delta(\theta-C) r_1 \cos(\psi_1 - \theta) \\
    \dot{\eta} &=  K \delta'(\theta-C) r_1 \sin(\psi_1 - \theta)  + K  \delta(\theta-C) r_1 \cos(\psi_1 - \theta)
\end{align}
Seeking the discrete spectrum first, we get
\begin{align}
    \lambda b(\theta) &=  K \delta'(\theta-C) r_1 \sin(\psi_1 - \theta)  + K  \delta(\theta-C) r_1 \cos(\psi_1 - \theta)
\end{align}
And then we invoked self consistency,
\begin{align}
    r_1 e^{i \psi_1} &=  \int e^{i \theta'} b(\theta') d \theta' \\
    \lambda r_1 e^{i \psi_1} &=  K r_1 \sin(C-\psi)(-i \cos(C) + \sin(C)) \\
    e^{i \psi_1} &=  K \sin(C-\psi)(-i \cos(C) + \sin(C)) 
\end{align}
Separating real and imaginary parts yields
\begin{align}
    \lambda \cos(\psi_1) &= K \sin C \sin (C - \psi_1) \\
    \lambda \sin(\psi_1) &= -K \cos C \sin( C - \psi_1) \label{temp}
\end{align}
First solve for $C$ by dividing the second by the first equations
\begin{align}
    \tan \psi_1 &= -1 / \tan C \\
    \tan \psi_1 \tan C &= -1
\end{align}
Which has solution $C = \psi_1 - \pi/2$. Subbing this into Eq.~\eqref{temp},
\begin{align}
    \lambda \sin(\psi_1) &= -K \cos (\psi_1 - \pi/2) \sin( \psi_1 - \pi/2 - \psi_1) \\
    \lambda \sin(\psi_1) &= - K \cos (\psi_1 - \pi/2) \sin(- \pi/2) \\
    \lambda \sin(\psi_1) &= K \cos (\psi_1 - \pi/2) \\
    \lambda \sin(\psi_1) &= K \sin (\psi_1) \\
    \lambda  &= K 
\end{align}
So we get $\lambda = +K$ which is the wrong sign?